N° d'ordre : 332-2007                                    Année 2007

**THESE**

**présentée devant :**

**l'UNIVERSITE CLAUDE BERNARD - LYON 1;**

**L'INSTITUT DE BIOLOGIE MOLECULAIRE ET DE GENETIQUE DE**
**L'ACADEMIE DES SCIENCES D'UKRAINE;**

**pour l'obtention du**

**DIPLOME de DOCTORAT**
**Spécialité biotechnologie**
(arrêté du 25 avril 2002)
présentée et soutenue publiquement le : 20 décembre 2007
par :

# M. Andriy BEREZHETSKYY

<u>Titre</u> :

# Développement des biocapteurs conductométriques à base de phosphatases alcalines pour le contrôle de la qualité de l'eau

<u>Jury</u> :
Prof. N. JAFFREZIC-RENAULT, Présidente – France
Prof. J.-M. CHOVELON, Directeur de thèse – France
Dr. S. DZYADEVYCH, Directeur de thèse – Ukraine
Dr. C. GONDRAN, Rapporteur - France
Dr. A. KUKLA, Rapporteur– Ukraine
Prof. A. SOLDATKIN, Examinateur - Ukraine
Dr. C. DURRIEU, Examinateur – France
Prof. C. TRAN-MINH, Examinateur invité – France

# UNIVERSITE CLAUDE BERNARD - LYON I

| | |
|---|---|
| **Président de l'Université** | **M. le Professeur L. COLLET** |
| Vice-Président du Conseil Scientifique | M. le Professeur J.F. MORNEX |
| Vice-Président du Conseil d'Administration | M. le Professeur R. GARRONE |
| Vice-Président du Conseil des Etudes et de la Vie Universitaire | M. le Professeur G. ANNAT |
| Secrétaire Général | M. J.P. BONHOTAL |

# SECTEUR SANTE

## *Composantes*

| | |
|---|---|
| UFR de Médecine Lyon R.T.H. Laënnec | Directeur : M. le Professeur D. VITAL-DURAND |
| UFR de Médecine Lyon Grange-Blanche | Directeur : M. le Professeur X. MARTIN |
| UFR de Médecine Lyon-Nord | Directeur : M. le Professeur F. MAUGUIERE |
| UFR de Médecine Lyon-Sud | Directeur : M. le Professeur F.N. GILLY |
| UFR d'Odontologie | Directeur : M. O. ROBIN |
| Institut des Sciences Pharmaceutiques et Biologiques | Directeur : M. le Professeur F. LOCHER |
| Institut Techniques de Réadaptation | Directeur : M. le Professeur L. COLLET |
| Département de Formation et Centre de Recherche en Biologie Humaine | Directeur : M. le Professeur P. FARGE<br>Directrice : Mme. le Professeur M. HEYDE |
| Département de Production et Réalisation Assistance Conseil en Technologie pour l'Education | |

# SECTEUR SCIENCES

## *Composantes*

| | |
|---|---|
| UFR de Physique | Directeur : M. le Professeur A. HOAREAU |
| UFR de Biologie | Directeur : M. le Professeur H. PINON |
| UFR de Mécanique | Directeur : M. le Professeur H. BEN HADID |
| UFR de Génie Electrique et des Procédés | Directeur : M. le Professeur A. BRIGUET |
| UFR Sciences de la Terre | Directeur : M. le Professeur P. HANTZPERGUE |
| UFR de Mathématiques | Directeur : M. le Professeur M. CHAMARIE |
| UFR d'Informatique | Directeur : M. le Professeur M. EGEA |
| UFR de Chimie Biochimie | Directeur : M. le Professeur J.P. SCHARFF |
| UFR STAPS | Directeur : M. le Professeur R. MASSARELLI |
| Observatoire de Lyon | Directeur : M. le Professeur R. BACON |
| Institut des Sciences et des Techniques de l'Ingénieur de Lyon | Directeur : M. le Professeur J. LIETO |
| IUT A | Directeur : M. le Professeur M. C. COULET |
| IUT B | Directeur : M. le Professeur R. LAMARTINE |
| Institut de Science Financière et d'Assurances | Directeur : M. le Professeur J.C. AUGROS |

**Acknowledgments**

First of all, I wish to thank the French Embassy in Ukraine and Rhône-Alpes region for their financial support.

The present work was carried out in the following establishments:

-Laboratory of Biomolecular Electronics (LBE) of Institute of Molecular Biology and Genetics (IMBG) of National Academy of Sciences of Ukraine (NASU);

-Institut des Recherches sur la Catalyse et l'Environnement de Lyon (IRCELYON);

-Centre SPIN (SPIN) de l'Ecole Nationale Supérieure de Mines de Saint Etienne (EMSE);

-Laboratoire des Sciences de l'Environnement (LSE) de l'Ecole Nationale des Travaux Publics de l'Etat (ENTPE).

These institutions are gratefully acknowledged.

I wish to thank my supervisors: Doctor Sergei Dzyadevych of LBE IMBG NASU, Professor Jean-Marc Chovelon of IRCELYON, Professor Canh Tran-Minh of SPIN EMSE and Doctors Claude Durrieu and Yves Perrodin of LSE ENTPE for accepting me in their establishments for my "co-tutelle" thesis. I am infinitely grateful for their confidence in me, their priceless help and generosity and, of course, for our fruitful scientific discussions during these years.

I warmly thank Academician Anna V. El'skaya, the Director of IMBG NASU, for creation the LBE and giving me the chance to start my research work in the this team.

I am also very grateful to Doctor Chantal Gondran and Doctor Alexander Kukla for accepting to be official reporters of this work and for their opinions.

I wish to thank Professor Nicole Jaffrezic-Renault and Professor Alexey Soldatkin for accepting to be members of my PhD thesis jury and for their sympathy and moral support.

I thank Hanh Nguyen-Ngoc from the University of Technology HCM, Ho Chi Minh, Vietnam, for initiating me in sol-gel technologies and for our fruitful discussions.

I would like to thank Houssemeddine Guedri for his gentleness and kind help with some manipulations.

My special thanks to Anne-Marie Danna of SPIN EMSE as well as to Therese, Isabelle, Alicia, Marc Danjean of LSE ENTPE for their priceless technical assistance in laboratory and for precious help in my "démarches administratives".

I want to warmly thank all my colleagues and friends for the unforgettable time we have had together:

i) David, Clotilde, Cecile, Guru, Alain, Sylvie, Helene Larmet, Helene Delhaye, Emilie, Murriel, Joseph Pollaco, Jose Soria, Jean-Phi, Laurence, Tierri, Marc, Urbain, Ruth, Anne-Laure, Rafael and all others from LSE ENTPE;

ii) Irina, Alena, Luda, Larissa, Sasha, Vika, Valentina Arhipova, Olga A Biloivan, Yaroslav Y Korpan, Nina R Polischuk, Alexandr E Rachkov and all others from LBE IMBG NASU.

I am inexpressible grateful to Elena Iv. Martinenko, Olga Bruyaka, Oleg Schuvailo and Stephan Collignon for their kindness, patience and for every possible help given me during all the time of our friendship.

Finally, I would like to thank my parents and my wife Olen'ka Sosovska for their constant and unconditional support: nothing would have been possible without it.

# Publications

**Articles in peer reviewed journals:**

1. V. M. Arkhypova, **A. L. Bereghetskyy**, O. A. Shul'ga, J.-M. Chovelon, O. P. Soldatkin, S. V. Dzyadevych. Investigation and optimization of conductometric transducers based on planar technology, **Sensor Electronics & Microsystem Technologies**.-2005.- 2. p. 48-54.

2. O.F.Sosovska, **A.L. Berezhetskyy**. Development of conductometric biosensor based on alkaline phosphatase for cadmium ions determination, **Ukrains'kii Biokhimichnii Zhurnal**.- 2007.-79.-4.p. 102-109.

3. **A.L. Berezhetskyy**, C. Durrieu, H. Nguyen-Ngoc, J.-M. Chovelon, S.V. Dzyadevych, C. Tran-Minh. Conductometric biosensor based on whole-cell microalgae for heavy metal ions determination, **Biopolymers & Cell**.- 2007. V.23, No 6.- p. 518-511

4. **A. L. Berezhetskyy**, O. F. Sosovska, C. Durrieu, J.-M. Chovelon, S. V. Dzyadevych, C. Tran-Minh. Alkaline phosphatase conductometric biosensor for heavy-metal ions determination**, ITBM-RBM**.-2008 in press. doi:10.1016/j.rbmret.2007.12.007

5. **A. L. Berezhetskyy**, O.F. Sosovska, J-M Chovelon and S .V. Dzyadevych. Development of conductometric biosensor based on alkaline phosphatase for heavy-metal ions determination, **Ecology and noospherology**.-submitted.

**Conférences :**

1. **A.L. Berezhetskyy**, C.Durrieu, J.-M.Chovelon, S.V.Dzyadevych, S.Cosnier, R.S.Marks, C.Tranh-Minh. Biocapteur conductometrique utilisant une matrice polypyrrole-alginate pour la détermination des métaux lourds, Xème colloque du groupe Française de Bioélectrochimie, Céret, France, 2006, Avril 5-7

2. **A. L. Berezhetskyy**, O. F. Sosovska, C. Tran-Minh, J.-M. Chovelon, S. V. Dzyadevych. Conductometric biosensor based on green microalgae *Clorella vulgaris*, 2nd International scientific and technical conference "Sensor Electronics and Microsystem Technologies" (SEMST-2). Odessa, Ukraine, June 26-30, 2006

3. O. F. Sosovska, **A. L. Berezhetskyy**, J.-M. Chovelon, A. P. Soldatkin, S. V. Dzyadevych. Conductometric biosensor based on alkaline phosphatase for heavy metal ions determination, 2nd International scientific and technical conference "Sensor Electronics and Microsystem Technologies" (SEMST-2). Odessa, Ukraine, June 26-30, 2006

# Table of contents

**General introduction**

Water is an indispensable resource for different human activities and for life in general. Water ecosystems are very fragile and delicate. To safe the quality of water resources a new European regulations was proposed for execution from 2015 [1].

Our research has been motivated by the importance of heavy metals determination. Heavy metals pollution is a menace for the environment and people's health. Heavy metal ions are toxic and non-biodegradable. They can be accumulated in the food chains. Therefore, there is a need to make the detection at such low concentrations as 1 ppb. In addition, to succeed objectives of the new regulations *Early Warning Systems (EWS)* for ecological monitoring have to be created. These systems must be able to work in *on-line* and *in-situ* modes.

Traditional means of heavy metals ions determination include: classic methods of analytical chemistry (atomic-absorption spectrophotometry, mass-spectrometry, HPLC), electrochemical methods (ex.: ion-selective electrodes and stripping voltammetry), toxicological approach of biotests uses inhibition effect to living organisms: trout, daphnia and bacteria (ex.: Microtox®, Truitel®). These techniques are expensive and/or need laboratory conditions so they cannot be used widely as field equipment.

Traditionally biosensors were used as biomedical instruments. Since 1980's they were adopted for environmental applications. In general, there are several advantages of biosensors for the monitoring of heavy metals in environment: time of analysis is sufficiently short for creation the EWS; sensors keep the sensitivity of classic toxicological tests, while being free from their inconveniences, thus field measurements are possible; relatively low price allowing multiplying the number of analyses.

At present, there are biosensors based on some enzymes, which partially meet a requirement of the EWS conception. They are founded on inhibition of an enzyme on the surface of biosensor transducer in the presence of a pollutant. For example, with urease based biosensor it is possible to measure concentration of heavy metal ions with detection limit about 1 ppm [2]. These detection limits are insufficient for majority of environmental monitoring needs. Thus, utilization of more sensitive enzymes is indispensable. There is very limited number of enzymes for this type of application. They are very labile and/or too expensive in purified forms. Thereby, this type of biosensor was not commercialized.

To overcome this obstacle it is possible to use living cells, immobilized on transducer, as natural enzyme bioreactor. A living cell has different enzymes necessary for its metabolism. These enzymes can be a specific target for certain pollutant families. Thus, the estimation of enzyme activity changes can help us to identify the presence of some xenobiotic species in the sample.

It is well known that microorganism's growth can be inhibited by most of pollutants families, whereas an individual enzyme specifically inhibits one contaminant class (ex. alkaline phosphatase can be specifically inhibited by heavy metal ions, esterase can be specifically inhibited by organophosphoric pesticides etc). Such pioneer approach in environmental biosensors presents the advantages of classic toxicological tests, but it is free from their inconveniences. Thus field measurements and utilization as a part of EWS are possible.

It should be noted, there are some whole cell biosensors. The biosensors based on genetically modified organisms are too much sophisticated for field application. Also the systems based on some principles like fluorescence are to much complicate for mass production.

In recent times, a French-Ukrainian research project was launched which was focused on conductometric microsensors with immobilized green algae *Chlorella vulgaris* for the detection of some pollutants. This project constituted a continuation of the PhD of Céline Chouteau [3] through which a collaboration between LACE (today IRCELYON) [4] and Laboratoire des  Sciences de l'Environnement de l'Ecole Nationale des Travaux Publics de l'Etat [5] was initiated. Among the advantages of such biosensors there is very low price because of mass produced microsensors and easy cultivated microalgae. By that means the biosensors will give the opportunity to multiply number of analysis for the better environmental monitoring. Some results have already been obtained by our team. It was demonstrated a conductometric biosensors based on microalgal alkaline phosphatase activity for detection of Cd, Zn and Pb with challenging detection limits [6, 7]. In these works immobilization of algae was performed by cross-linking of cells with albumin molecules in vapors of glutaraldehyde. This method has a number of inconveniencies: lost of enzymatic activity of algae because of glutaraldehyde toxicity and back side reactions in the biosensor membrane between analytes and bovine serum albumin [3]. Thus, it is necessary to find better immobilization method.

Present work is devoted to *development of an alkaline phosphatase based conductometric biosensors for an assessment of heavy metal ions in water.*

Main *aims* of this work are following:

- to investigate thin-film planar conductometric transducer characteristics depending on electrodes material, dimensions and geometry

- creation of conductometric biosensors based on alkaline phosphatases of different sources sensitive to heavy metal ions;

- to resolve the problem of physicochemical instability and back-side reactions of immobilizing system;

- application of proposed biosensor to estimate heavy metals toxicity of  urban water.

This work is an offspring of collaboration between:

-Ecole Nationale Supérieure de Mines de Saint Etienne [8];

- Laboratoire des Sciences de l'Environnement de l'Ecole Nationale des Travaux Publics de l'Etat [5];

-Institut des Recherches sur la Catalyse et l'Environnement de Lyon (IRCELYON) [4];

-Laboratory of Biomolecular Electronics of Institute of Molecular Biology and Genetics of National Academy of Sciences of Ukraine [9].

The *originality* of this work is utilization of sol-gel supports to combine high sensitive to heavy metals alkaline phosphatases with challenging thin-film planar conductometric transducers. This combination of traditional biosensor parts will give the possibility to detect very low heavy metal concentrations in the environmental samples.

The manuscript includes: a first chapter is bibliographic review illuminating conductometric biosensors and environmental monitoring problem; a second chapter described the fundamentals of conductometric transducers and optimization of thin-film planar electrodes for biosensor applications; a third chapter examining the possibility to create and to optimize conductometric biosensor based on bovine alkaline phosphatase for heavy metals ions detection; the fourth chapter devoted to creation and optimization of conductometric biosensor based on alkaline phosphatase active microalgae and sol gel technology; the last chapter described application of the proposed biosensor for measurements of heavy metal ions toxicity of urban water; general conclusions stating the progresses achieved in the field of environmental monitoring.

# Chapter 1

## Chapter 1.  Bibliographic review

### Résume en français du chapitre 1

Ce premier chapitre couvre la technologie relative aux biocapteurs conductométriques. Plus spécifiquement, il permet de :

(a) décrire les principes de la conductométrie ;

(b) faire une revue sur les transducteurs conductométriques, notamment ceux qui sont utilisés dans la bioélectronique ;

(b) faire une revue des méthodes classiques et récentes d'immobilisation des biomatériaux ;

(c) faire une revue des biocapteurs utilisés dans l'environnement.

A partir du concept des biocapteurs et de leurs classification (schéma ci-dessous), nous avons focalisé notre recherche sur des points bien précis (indiqués en rouge dans le schéma) : 1/ les membranes biosélectives catalytiques avec comme matériau des cellules entières ou des enzymes purifiées 2/ le capteur électrochimique avec dans notre cas des microélecrodes conductométriques et enfin 3/ la possibilité de détecter des analytes de type ions de métaux lourds.

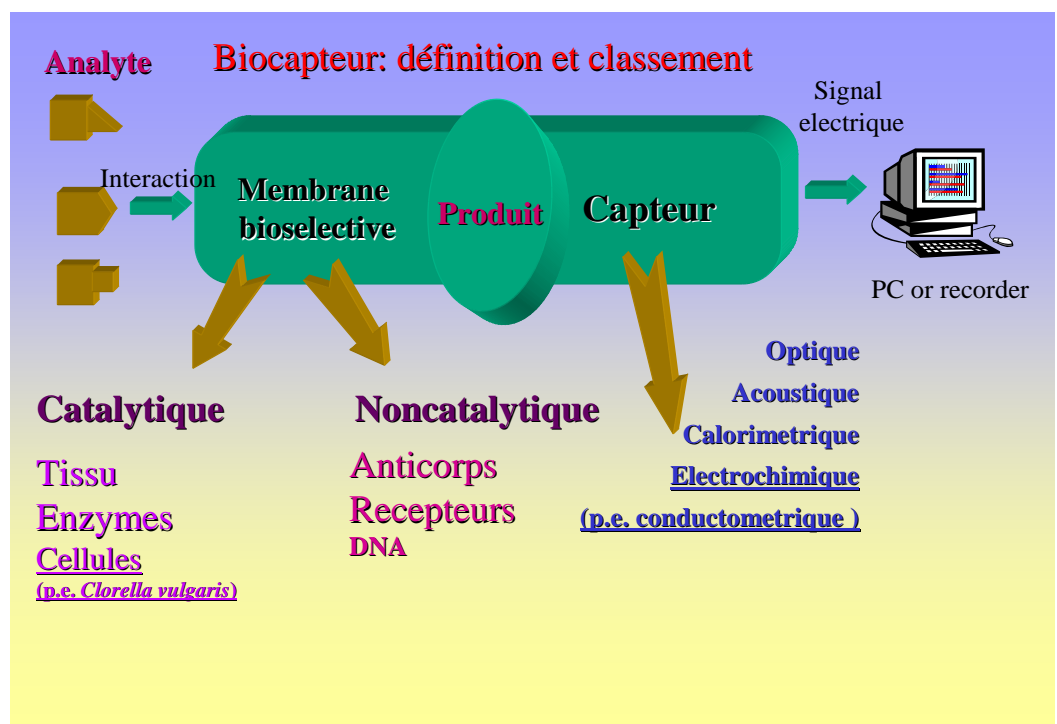

**Fig. 1.1**. *Définition et classement des biocapteurs [10]*

Dans ce chapitre une étude bibliographique sur les fondements des capteurs conductométriques a été réalisée en tenant compte de leurs applications potentielles dans le domaine de la surveillance environnementale. De même, pour la fabrication des biocapteurs, les différentes options résultantes de techniques récentes liées au développement des microsystèmes ont été analysées pour ce qui est de la fabrication des transducteurs, les méthodes d'immobilisation du biomatériel, la combinaison de plusieurs technologies dans un seul cycle de production.

Nous avons montré que les microtechnologies les plus avancées présentaient certaines particularités qui ont été utilisées pour la production des biocapteurs électrochimiques. Par ailleurs, les schémas des dispositifs pour les mesures expérimentales ont été rappelés. Nous avons également décrit dans les détails les acquis dans le domaine du développement des biocapteurs enzymatiques.

Au stade initial du développement des biocapteurs, le point clé repose sur le choix d'un transducteur approprié et le plus efficace possible en terme de caractéristiques analytiques et de procédure de mesure sur le terrain. Malheureusement à ce niveau, le choix optimal n'intègre pas, en générale, l'ensemble du processus analytique allant de la conversion du signal mesurée, aux propriétés spécifiques des transducteurs ou encore des matériaux biologiques. Donc, une étude compréhensive sur les interactions entre les éléments biologiques et les transducteurs physiques est extrêmement importante pour le développement d'une stratégie générale d'utilisation des transducteurs de différents types dans la bioélectronique en fonction du problème analytique à résoudre.

A ce jour ce sont les biocapteurs ampérométriques qui sont le plus souvent utilisés d'un point de vue commercial du fait d'études plus approfondies à leur égard et des avantages qu'ils peuvent offrir. En ce qui concerne les différents transducteurs conductométriques, ils doivent encore davantage être étudiés et évalués pour ce qui est de l'efficacité de l'enregistrement des signaux physiques et chimiques générés, à l'interaction entre le matériel biosélectif et le substrat analysé selon les conditions expérimentales, le matériel d'électrode et sa géométrie.

Il relève donc d'une priorité urgente, s'il s'agit de la commercialisation et du potentiel de capacité concurrentielle des biocapteurs conductométriques qui pourraient être favorisés par rapport à d'autres systèmes analytiques, d'améliorer la sélectivité, la sensibilité, la facilité opérationnelle et la rapidité.

Le domaine d'application principale des biocapteurs est le diagnostique médical où l'on utilise de nombreux outils commerciaux. Cependant, l'application des biocapteurs dans l'environnement reste insuffisante et par conséquent, cela doit être considéré comme un défi important. Les outils basés sur les multicapteurs et l'utilisation de multienzymes sont prometteurs, plus particulièrement, dans le cas d'utilisation des techniques en microélectroniques avancées.

## 1.1 Introduction

The requirements and regulations in the fields of environmental protection, control of biotechnological processes, and certification of food and water quality are becoming more and more urgent. At the same time stricter requirements regarding human and animal health have led to a rising number of clinical and veterinary tests. Therefore, there is a need in developing highly sensitive, fast and economic methods of analysis. The elaboration of biosensors is probably one of the most promising ways to solve some problems concerning sensitive, fast, repetitive, and cheap measurements [11].

A biosensor converts the modification of the physical or chemical properties of a biomatrix, which occurs as a result of biochemical interactions, into an electric or an optic signal whose amplitude depends on the concentration of defined analytes in the solution. Functionally, the device consists of two parts: a biomatrix, *i.e.* a detecting layer of immobilized material (enzymes, antibodies , receptors, organelles, microorganisms, and a transducer (potentiometric, impediometric, amperometric, conductometric, acoustic, optic or colorimetric) [12].

Although numerous reviews, books as well as a lot of experimental research concerning various types of biosensors have been published. There are some unsolved problems associated with MEMS technology: electrode manufacturing, methods of immobilization of biological material and combination of different processes into a common technological cycle [13-15].

Generally, microsystems are of micron- or millimeter size, the lowest limit being often even less, so, nanotechnology and ultra microelectrodes with nanodimensions are typical microsystem elements as well. Thus, microsystem technologies can be referred to as the technologies associated with microsystems and combining original heterogeneous technological procedures of manufacturing electronic, mechanical, optical, and other components on the basis of planar lithographic processes.

Conductometric (bio)sensors are a very promising class of analytical devices with high sensitivity. Almost all electrochemical analytical methods are based on electrode electrochemical reactions (potentiometry, voltamperometry, amperometry, and coulometry). Conductometry is a method in which either there are no electrochemical reactions on the electrodes at all or they are secondary and can be neglected. Therefore, in conductometric method the most important property of an electrolytic solution is its conductivity, which varies in accordance with quite a wide range of enzymatic reactions.

The liquids analyzed are mostly considered to have significant background conductivity which is easily modified by different factors, therefore, the selectivity of this method is presumed to be low and consequently its potential use for different application was thought doubtful. However, in case of integral microbiosensors, most of these difficulties can be overcome using a differential measuring scheme which compensates for changes in background conductivity, the influence of temperature variations and other factors.

### 1.2. Fundamentals of conductometric method of measurement.

As a rule, an alternative current (AC) is used in conductometric analysis. Therefore, at first it is necessary to describe the processes occurring in an electrochemical cell and at a metal-solution interface, under a sinusoidal AC. Then we will study the parameters influencing the solution conductivity in order to develop experimental measuring schemes.

### 1.2.1. Electrochemical impedance of the metal-solution system.

In order to understand the basic operating principle of conductometric transducer, the nature of surface impedance of the electrolyte-metal interface must be understood. The key research approach in this system is to interpret the physical and chemical processes like some equivalent electric circuit with electronic elements, capacitors and resistors, which simulate real processes.

Let us describe the electrode process at the molecular level.

The interaction between separate charged or polar particles takes place within a homogeneous phase, the total force affecting any particle being equal to zero. At equilibrium, cations and anions of an electrolyte, as well as molecules of water, are distributed homogeneously. Therefore, the solution is electrically neutral. In the vicinity of the phase interface, the equilibrium of the different forces affecting each particle is violated due to the difference in properties of the molecules placed at opposite sides of the interface (*e.g.* the molecules at the electrode surface and most of the electrolyte ions and dipole molecules at the opposite surface). This is the same for electrons and atoms of the electrode material. Therefore, right near the phase interface the total vector of the forces affecting each particle is not zero. As a result of this anisotropy, the particles placed at various distance from this surface are oriented or reoriented driven by these forces. In this alternative force field the particles move towards the lowest energy state. This process takes place at a distance of several ion radii and can cause more or less dipole orientation of the solution molecules in the vicinity of the electrode surface. The oriented dipoles on the electrode surface can be considered as a charged plate capacitor, *i.e.* a double electric layer occurs likewise in the capacitor. Thus, the quantitative meaning is introduced – the capacity of double electric layer $C_{dl}$, usually $C_{dl} = 10$-$20 \ \mu F/cm^2$ [16].

Some of the charged particles can pass through the double layer and cause electrochemical reactions on the electrode surface. This process, called chemical polarization, can be described by a defined resistance, called penetration resistance, $R_p$. It can be estimated from the Buttler-Wolmer equation and at low-amplitude applied signal can be expressed as follows:

$$R_p = \frac{RT}{nFi_o}$$ (1.1),

where R is ideal gas constant, T is absolute temperature, F is Faraday constant, n is number of electrons participating in the electrode reaction, $i_o$ is density of exchange current.

In the case of concentration polarization, a sharp decrease of the discharged ions concentration, in the presence of electric current, is soon revealed in the layer adjacent to the electrode, due to the low speed of supplying particles from bulk electrolyte towards the electrode surface by diffusion. This results in decreasing current and increasing electrode potential. Therefore, the phase shift between periodically changed current and electrode potential is revealed, the potential change always occurring after the current change. In such a system the concentration polarization due to the ion diffusion from the phase interface into bulk electrolyte causes an increase in the surface impedance, particularly at low frequencies.

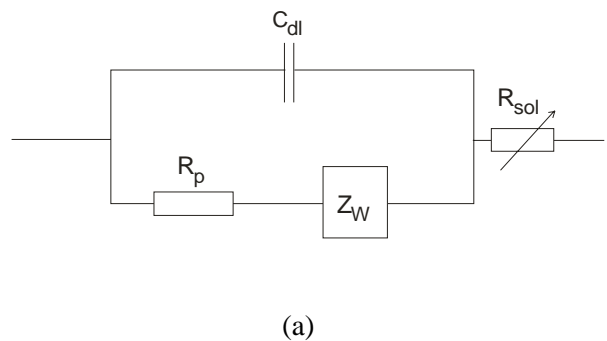

(a)

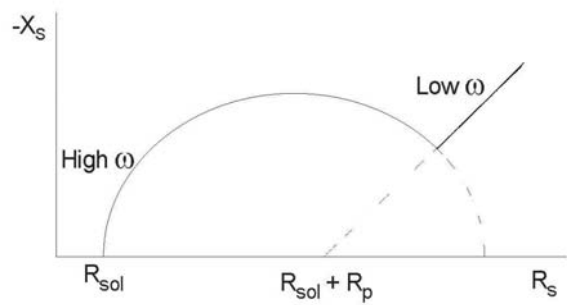

(b)

**Fig. 1.2.** *Classic equivalent circuit (a) and corresponding impedance curve (b) for the metal-electrolyte interface [17]*

This diffusion, or "Warburg" impedance $Z_w$ depends on frequency and is described by equation:

$$Z_w = \frac{RT}{n^2F^2}(D\omega)^{-1/2} \tag{1.2},$$

where $\omega$ is angular frequency, $D$ is diffusion coefficient.

Therefore the electrochemical impedance of the "metal electrode – solution" system can be simulated by the equivalent scheme as follows (Fig. 1.2 a), where $C_{dl}$ is double layer capacity, independent of frequency; $R_p$ is polarization resistance simulating chemical polarization and, like $C_{dl}$, independent of frequency; $Z_w$ is diffusion impedance simulating concentration polarization and depending on frequency; $R_{sol}$ is electrolyte resistance. The corresponding impedance curve is presented in Fig. 1.2 b. At a frequency of measurement higher than 10 Hz the diffusion impedance can be neglected and the total system impedance consists mainly of the solution resistance, the double layer capacity, and the penetration resistance.

The interface impedance depends on the solution's nature and composition and frequency in different way. Therefore, depending on the kind of analytical signal used for high- and low-frequency conductometry, the physicochemical analysis based on the measurements of interface impedance, respectively, or total electrochemical cell impedance, can be distinguished [17].

### 1.2.2. Electric conductivity of solutions.

The conductivity of liquids results from dissociation of the dissolved substance, electrolyte, into ions and migration of the latter induced by an electric field. In the presence of electric current, electrolytes dissociate into atoms or atom groups which are parts of the solvent molecules, namely, ions, *i.e.* the electrolyte conductivity has an ion character [18].

When the potential is applied to the electrode, there is an electric field within the electrolyte, so the chaotic ion movement is influenced by the ordered, oppositely directed movement of ions (those with negative charge move towards anode, while positively charged ones move towards cathode). Thus, the current in the electrolyte is caused by the ion movement towards the electrodes where the ions are neutralized and isolated as neutral atoms (or molecules).

The ion flux, *i.e.* number of ions passing through a unit of electrolyte cross-section per unit of time ($p_i$) can be determined by the formula:

$$p_i = c_i \, v - k_i \, c_i \, \text{grad} \, \mu_i - z_i \, v_i \, c_i \, F \, \text{grad} \, \psi \qquad (1.3)$$

where $v$ is speed of solution flow due to natural or forced convection; $c_i$ is ion concentration; $k_i$ is diffusion coefficient; $z_i$ is charge number; $v_i$ is speed of ion movement caused by applied field; $F$ is Faraday number.

Thus, the first member in formula (1.3) corresponds to the contribution of the convectional flow of ions at a concentration of $c_i$, the second – the contribution of their molecular diffusion, the third – that of ion migration induced by the supplied potential. Temperature, as a rule, is assumed to be constant (T = const, grad T = 0), so the ion thermal diffusion can be ignored.

In reality all three processes usually coexist, but have an initial influence on each other. In any case the assumption can be made that only one of them is taken into account. Thus, in case of homogenous immobile electrolyte, the first and second members of equation (*1.3*) can be neglected, and only the ion migration caused by electric field effect be considered. Then

$$p_i = - z_i \, v_i \, c_i \, F \, \text{grad} \, \psi = - c_i \, u_i \, \text{grad} \, \psi \qquad (1.4)$$

where $v_i$ is speed of ion movement, $c_i$ is ion concentration, $u_i = z_i \, v_i \, F$ is ion mobility which is a constant value for the given ion in the infinitely dissolved solution.

The current density, *i.e.* the current per unit of the system cross-section is an algebraic sum of products of ion fluxes and ion charges:

$$j = F \sum z_i \, p_i = F \, \text{grad} \, \psi \sum z_i \, c_i \, u_i \qquad (1.5)$$

On the other hand, according to the Ohm's law

$$j = S \, \text{grad} \, \psi \qquad (1.6.)$$

where $S$ is conductivity, *i.e.* value reciprocal to resistance.

Hence, from (*1.5*) and (*1.6*)

$$S = F \sum z_i \, c_i \, u_i \qquad (1.7)$$

Thus, the conductivity of electrolyte solution depends on the ion concentration and mobility.

The resistance of electrolyte solution is well-known to be in direct proportion to the distance "L" between the immersed electrodes and reciprocal to their area A, therefore

$$S = \chi \frac{A}{L} \qquad\qquad (1.8),$$

where $\chi$ is specific conductivity.

This leads to the following conclusions. Conductometric measurement commonly consists of determining the conductivity of the solution between two parallel electrodes; its value is a sum of conductivities contributed by all ions within the solution tested. Biospecific reactions can cause occurrence of new ions, as well as changes in ion concentration and mobility. This results in changing solution conductivity registered by a conductometric transducer.

### 1.2.3. Conductometric measuring schemes.

It has been stated that conductometric analysis is based on the measurement of conductivity of an electrolyte solution. An active resistance between the electrodes immersed in the tested solution is determined. The bridge circuit for measuring electric resistance was first developed by Wheatstone, while Kolrausch has used it for alternative current (Fig. 1.3 a). This bridge circuit consists of four resistances: $R_x$ is resistance of the analyzed solution, $R_m$ is set of resistances, $C_m$ is set of capacities, $R_1$, $R_2$ are definite resistances ($R_1 = R_2$), n is zero-indicator.

According to the bridge circuit theory, the circuit has the highest sensitivity when the resistances of all four arms are equal. Since as a rule $R_1 = R_2$, the AC bridge circuit balance can be reached only when pure resistances and reactances taken separately are balanced. The minimum zero-indicator value can be obtained by varying active resistances and capacitances. Then the $R_m$ value corresponds to the value $R_x$ of the resistance between electrodes. The resistance of electrolyte solution can be determined with an accuracy of 0.5 – 1% by means of such a bridge circuit.

However, the bridge circuit of measurement supposes constant additional manipulations with a box of active resistances and capacitances that decrease its accuracy. Fig. 1.3 b demonstrates the modified Wheatstone bridge circuit which enables direct registration of AC in the cell tested without any subsidiary steps.

The measuring scheme shown in Fig. 1.3c uses an operational multiplicator of high resistance. In this case, conductivity $S_x$ is in direct proportion to the value of current at the operational multiplier output. This scheme has been used as a basis in the research of enzyme conductometric biosensors [19].

The 4-electrode scheme is also employed. In it, an AC current is applied to one electrode pair while the voltage drop is measured on the other electrode pair which is not polarized by the current and thus serves as a probe.

The schemes described are quite simple and mostly used in experiments. Each of them has its own advantages and disadvantages; therefore the choice has to be made for each specific case.

### 1.3. Conductometry in enzyme catalysis.

The conductometric measuring method can be used in enzyme catalysis to determine substance concentration, and the enzyme activity, selectivity in this case being provided by the enzymes which catalyze only certain reactions. As a matter of fact, the subject under consideration is not a biosensor as such but an application of this method in enzymology.

In 1961 one of the first research in this field was published showing how it might be possible to determine urea concentration in solutions [20].

This method is based on the difference between electric conductivity of urea solution and that of a solution of ammonium carbonate formed as a result of urea hydrolysis by urease. In the experiments a bridge measuring scheme was used. The urease activity was shown to decrease in the presence of heavy metal (Ag, Hg, *etc.*) ions in the solution. Such electrolytes as NaCl and KCl do not influence urease activity, but if their concentration in the solution is high it can lead to a wrong result, especially at low urea concentrations. At low electrolyte concentrations in experiments without buffer solution, during urea hydrolysis the medium pH gradually changed from 7.0 to 9.0. However, this causes only an insignificant change in the urease activity while the solution conductivity during the reaction varied substantially. The urea concentration was determined within the 0.1 $\mu$M – 2 mM range, at optimal pH 7.0. A comparison of the conductometric method with other methods of urea analysis carried out in that work has shown that the former is characterized by high accuracy, speed and simplicity. Besides, in contrast to optical methods, the measurement accuracy of conductometry does not depend on solution color.

In 1965 a paper was published on applying the conductometric method to study the kinetics of urea enzyme hydrolysis as well as to urease activity determination [21]. A differential measuring scheme was used in their experiments.

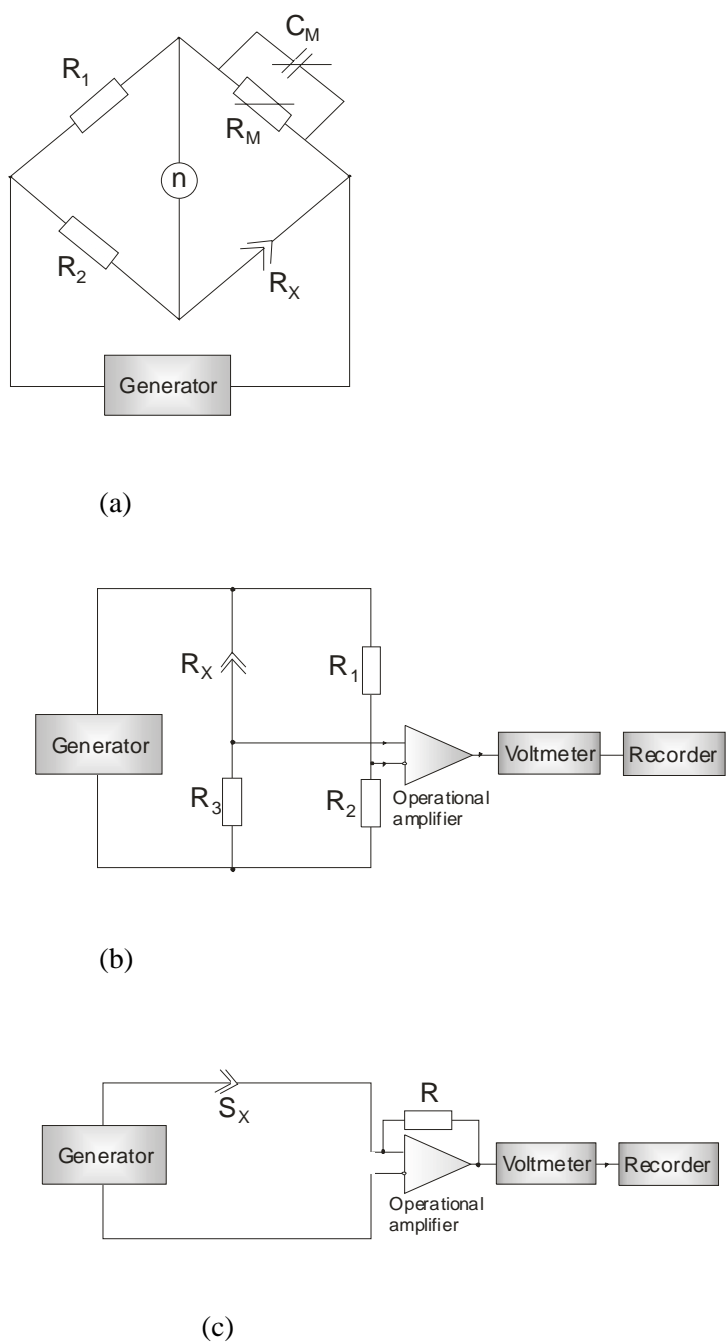

(a)

(b)

(c)

**Fig. 1.3.** *Schemes of conductometric set-ups  [19]*

The system consisted of two pairs of platinum plates, each of them placed in its own measuring cell, one with the enzyme, another without one. The difference between signals from both cells was registered, thus eliminating any error associated with variations of the parameters outside the cells (temperature, buffer concentration, *etc*.). The determination range of urea concentration was 1 – 75 mM, while that of urease activity was 0.04 –2.5 activity units/ml. Comparing the data obtained with the results of classic photometric analysis showed that the conductometric method has all merits of the classic one, and exceeds the latter in accuracy and speed.

At the same time in [22] a research work was presented demonstrating that the changes in conductivity during an enzyme reaction can be considered to be a universal characteristic of the substrate chemical transformation. Even if the conductivity of the products of the reaction and that of the substrate differs a little, the change in the solution viscosity and in the level of hydration of molecules and ions at the substrate transformation (especially in the presence of other current carriers in the solution) cause noticeable variation in the tested mixture. To prove the potentials of conductometry experimentally, the authors chose the reactions associated with different mechanisms of conductivity change. It was important to expose the character of conductivity change in case of evident changes in the solution composition as well as when only solution viscosity and level of molecule hydration vary as a result of the reaction. The enzyme hydrolysis of acetylcholine and starch, on the one hand, and the enzyme depolymerization of gialuronate, on the other, were considered to be the reactions meeting these requirements.

Acetylcholine hydrolysis is accompanied by the rupture of complex ether bound and the formation of acetic acid dissociating in protons and $CH_3COO^-$. The proton does not participate in the total conductivity mechanism because the reaction takes place in a phosphate buffer, pH 7.8, while the appearance of $CH_3COO^-$ anions in the solution results in an increase of the solution conductivity.

In case of starch hydrolysis, filtered human saliva was added to the starch solution to hydrolyze the latter and to decrease the solution viscosity, *i.e.* to increase the conductivity of solution. When gialuronidase was used, the conductivity increased as a result of reducing the viscosity of the solution due to the depolymerization of gialuronate molecules.

The authors have demonstrated that the conductometric method is preferable to the well-known routine methods of biochemical analysis because of its higher accuracy and lower labor consumption. Using a single device and a single method with no modification, the kinetics of three enzyme processes with specific features was studied, while three devices based on different principles would be required for classic enzyme analysis.

In the next work, the authors modified their conductometric device and used a differential mode to measure the activities of collagenase, trypsin, lactate dehydrogenase and pseudocholinesterase [23].

However, despite demonstrating the potential of the conductometric method to record enzyme processes, the character of all the works mentioned is that of a preliminary or feasibility study.

At the beginning of the 80s a detailed analysis concerning the potentials and limitation of the conductometric method of measurement was performed for the complex study of urease [24]. In the first part of the work, the effect on the conductivity of the medium pH, the urea, urease and salt concentrations of the studied solution was investigated in the absence of any enzyme reaction. It was shown that the changes in concentrations of the solution compounds cause no substantial variations of its conductivity which is why these changes can be neglected. The curve of conductivity–pH dependence is bell-shaped with a maximum at pH 8.0 for Tris-buffer and at pH 6.0 for a citrate buffer.

In the second part of the work the influence of the medium conditions mentioned on urea hydrolysis was investigated. Though pH 7.2 was shown to be optimal for urease in citrate buffer, the substantial dependence of the reaction rate on the solution ion strength was revealed at this value, the rate decreased with rising ion strength, while at pH 6.5 it hardly changed. The Michaelis' constant for urease in citrate buffer, pH 4.5, was determined to be around 2.5 mM. The linear dependence of the rate of the enzyme reaction on the urease concentration in the solution was also obtained. The successful use of the conductometric method to study the enzyme activity of urease, described in this work, can be considered as a convincing argument of its potentialities due to high sensitivity and good agreement of the kinetic parameters obtained by the conductometric method compared with the results of classic biochemical analysis.

In reference [25] a 6-channel conductometer is used to study different enzymes. The investigation allowed the authors to formulate 5 factors that bring about, separately or in combination with each other, a change in conductivity, thus ensuring the potential of the conductometric method to record parameters during enzyme reactions

In the report [26], peptide pyroglutamyl was studied by the conductometric method, the Michaelis constants were obtained as 0.34 mM and 0.47 mM for derivatives of alanine and tyrosine, respectively.

However, conductometric methods have some limitations. The ratio between the signal and noise level should not be lower than 2%. For this reason, the concentrations of buffer and some other ingredients, which can be added to the reaction mixture, are important. The method sensitivity is reduced in the presence of non-reacting ions in the solution. Buffers with low ion strength can be used, though, to measure low concentration until the signal/noise ratio is of proper value. A disadvantage of conductometry is also its low specificity – it is incapable of distinguishing between simultaneous reactions that can cause an artifact. The capacity of the double layer and the electrode polarization during reaction can be also sources of the method error. All the investigations reported have become a basis for further development of conductometric biosensors.

### 1.4. Novel materials and methods of biomaterial immobilization

The biomaterial immobilization can be determined as an involvement of biomolecules into an isolated phase separated from a solution phase but exposed to exchange molecules of substrate, effector and inhibitor with the latter.

Immobilized enzymes have some significant advantages as compared with native enzymes:

✓ A heterogeneous catalyst can be easily separated from the reactive medium which allows stopping the reaction at the definite moment, using the catalyst repeatedly, and obtaining the pure product unpolluted with the enzyme.

✓ Immobilized catalysts enable continuous enzymatic process, *e.g.* in a flow-through cell, as well as control over both rate of catalyzed reaction and product yield by changing flow rate.

✓ Enzyme immobilization allows regulating the enzyme activity, varying carrier properties as a function of some physical influence.

✓ Enzyme immobilization or/and modification promotes intended changes in catalyst properties, including specificity and dependence of catalytic activity on pH, ion composition and other parameters of environment.

To be or not to be a success in practical application of immobilized enzymes depends to a great extent on correct selection of a carrier. At present, numerous carriers, both organic and inorganic, are used for biomaterial immobilization. The carriers have to meet the following key requirements:

- ✓ High chemical and biological stability.
- ✓ High mechanical strength.
- ✓ Sufficient enzyme and substrate permeability, large specific surface, high capacity and porosity.
- ✓ Possibility to be produced in technologically suitable forms.
- ✓ Easy transition into reactive form (activation).
- ✓ High hydrophilic properties to ensure binding of enzyme with carrier in aqueous medium.
- ✓ Low cost.

The materials mostly used at immobilization are multifunctional agents (*e.g.* glutaraldehyde and hexamethyl diizocyanite) generating covalent bonds between biocatalytic particles or proteins and non-conductive polymers (*e.g.* polyacrylamide and polyphenol) forming a kind of a net capable of capturing and holding biomaterial.

An advantage of organic conductive materials is that as the biological object immobilized at the electrode surface is formed which contacts electrically with metal or carbon conductor. This results in a tighter connection between the enzyme redox-centre and the electrode surface. Such kind of polymers can be produced by various chemical processes, namely: the Zigler-Natta reaction for polyacetylene, metalorganic component coupling (polythiophene), monomer oxidation, *etc* [27].

To capture oxydoreductases the redox-polymer hydrogels were used based on binding polyethylene glycol, diacrylate and vinyl ferrocene using UV cross-linking. The enzyme was dissolved in the mixture directly before irradiation. The latex particles were also used at molecule immobilization, the latex two-dimensional membrane formation on the conductive solid surface was studied at immobilization of human serum albumin [28].

Various commercial polymers, such as polyvinyl chloride, polyethylene, polymethacrylate, NAFION[®] and polyurethane are utilized as subsidiary external membranes whose functions are diffusion control, mechanical protection and decrease of effect of interfering particles.

The main methods of enzyme immobilization are presented in Fig. 1.4

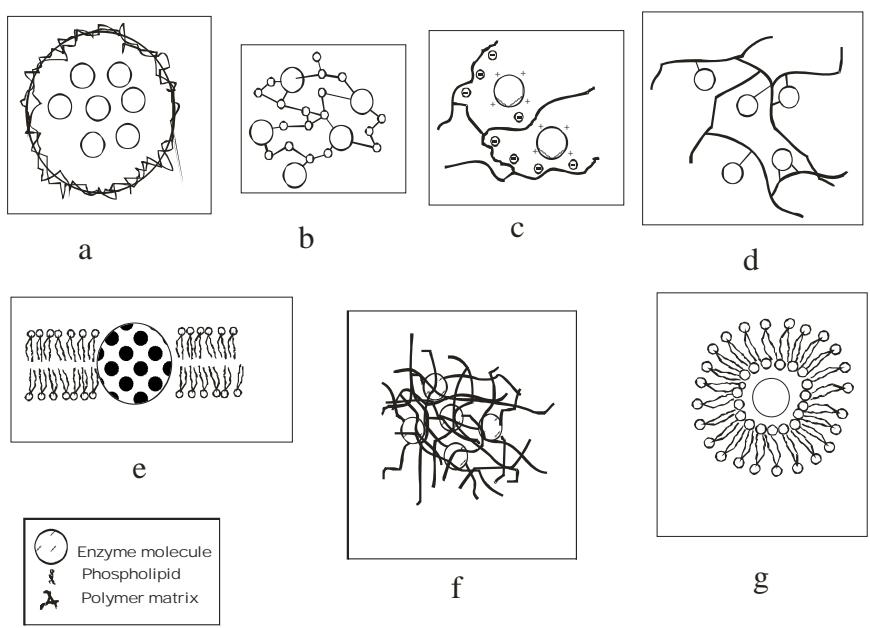

**Fig. 1.4.** *Possible methods of enzyme immobilization: a) encapsulation; b) co-polymerization; c) electrostatic binding; d) covalent binding; e) hydrophobic interaction; f) insertion into polymer; g) incorporation into liposome [10]*

First of all, the properties of immobilized enzyme depend on the phase into which it is involved. The enzyme molecule is often bound covalently to insoluble polymer, *e.g.* cellulose or polyacrylamide, in powder or film form. This method, though laborious, is mostly used since it assures tight binding of the immobilized enzymes with the polymer carrier. Sometimes enzyme molecules get covalently bound with one another or with some inert protein creating insoluble nevertheless active complex [29]. These methods are advantageous by at least two important characteristics: high strength of the conjugate formed and a possibility to vary substantially such enzyme features as substrate specificity, catalytic activity, and stability.

Another method of enzyme binding to polymer consists in utilization of electrostatic or other non-covalent binding mechanisms [30], when it is not necessary for an enzyme to be conjugated with the polymer, it can be inside the latter. In this case the polymer forms a net-like matrix around the enzyme with cells as little as to prevent a rather large enzyme molecule from getting free out of the net, and simultaneously big enough to be permeable for low-molecular weight substrates.

One of the encapsulation variants utilizes phospholipids bilayer as an enzymatic membrane phase [31]. The enzyme can either be barriered by phospholipids within the aqueous solution or dissolved directly within the hydrophobic part of phospholipid bilayer. To optimize the enzyme loading in this system, the alternate deposition is often used. In [32], for example, up to ten layers with cholinesterase and two additional layers with cholinesterase were deposited to increase the sensor sensitivity.

For conductometric biosensors an enzyme was included inside the film of an albumin gel by means of glutaraldehyde [19, 33, 34]. For example, in [19] urease was immobilized on the electrode surface by deposition of 10 μl of mixture consisting of the enzyme (100 mg/ml), BSA (100 mg/ml), and 2.5 % solution of glutaraldehyde at a temperature of 20°C; the 1.5 mm thick film of enzyme-albumin gel was formed in 9-10 min. In [34] enzymes were immobilized in the glutaraldehyde saturated vapors. A drop (0.1 μl) of the mixture of enzyme (50 mg/ml), BSA (50 mg/ml), and 10 % glycerol was deposited on sensitive surface of the transducer and then placed in the glutaraldehyde saturated vapors. In [35] a procedure of covalent binding of urease with the collagen membrane is reported.

Enzyme electrochemical polymerization onto the electrode surface in the presence of polymer is a novelty in enzyme immobilization techniques [36]. The enzyme is simply captured into the polymer matrix-carrier during electro polymerization. Electrochemical polymerization is technologically simple, allows selecting and maintaining the membrane dimensions, shape and thickness, ensures controlled precipitation, gives a possibility to manufacture different microbiosensors in one technological cycle [37].

Sol-gel technique is a perspective nanotechnology for biomaterial immobilization. Conventional glass-making processes require reaction at elevated temperatures which are not compatible with entrapment of fragile biomolecules. The sol–gel process involves the transition of a system from a molecular precursor state through the formation of nano-sized bricks during the sol colloidal phase into a solid gel phase and finally transitions into a dried ceramic material. Silica matrixes are relatively inexpensive to synthesize and have interesting properties including optical transparency, biocompatibility and chemical inertness. The design flexibility of sol–gel technique and simplicity of fabrication can fulfill to create the surfaces with structural and chemical features that could be compatible with biomaterials.

Immobilization of chemicals in such inorganic structures is not restricted to organic molecules. Since the sol–gel process can be carried out in moderate conditions, the structural elements of the nano-sized sol bricks are preserved and enable the encapsulation of biomolecules [38] and even bacteria [39]. Biomolecules entrapped in sol–gel matrixes typically exhibit improved resistance to thermal and chemical denaturation, and increased storage and operational stability. This technique has been applied to immobilize various biological molecules (proteins, enzymes, antibodies) using the alkoxide route with tetraethyl orthosilicate (TEOS) or tetramethyl orthosilicate (TEMOS) as precursors [40], but they are not really cell compatible because the by-products ethanol or methanol are harmful for the biological species [36]. These alcohols can be removed from the acidic aqueous sol prior gelation [41] or the alkoxides can be replaced by aqueous precursors [42].

Accurate option of the method of biomaterial immobilization is of great significance. The alternative is determined by a number of factors, some of which are indefinable until the method is tested. Previous empirical selection is usually performed to decide whether the specific carrier is necessary or any material is suitable to guarantee high activity of the immobilized enzyme. It is clear, that in the first case the search boundaries are constricted while in the second one the further study has to clarify whether or not the immobilization causes the enzyme inactivation and whether the immobilized enzyme is functional under necessary conditions. To ensure required immobilization, the following factors are to be taken into consideration:

✓ The enzyme stability both in the catalyzed reaction and during immobilization. The reactions which require the presence of strong basic or acidic compounds, organic dissolvents or temperature above 50°C are to be avoided.

✓    Covalent binding is desirable (if possible) to be performed by the cross-linking reagents interacting mainly with the chemical groups absent in the enzyme active centre. Since it is regularly unrealizable, the cross-linking reagents are to be as large as not to penetrate to the active centre. For example, the polymer "activated" cellulose is preferable in comparison with a small bifunctional reagent glutaraldehyde.

✓    The enzyme active centre should be protected, wherever possible, from the influence of immobilizing agents. For example, sulfohydrile enzymes can be treated with glutathione or cysteine prior to immobilization. Sometimes the inactive form of some enzymes, *e.g.* of papain, can be used for immobilization, at other times an active centre can be blocked by adding the substrate of enzyme at saturating concentration to the reactive mixture.

✓    The removal of non-linked enzyme by washing should not influence the immobilized enzyme negatively. For example, the cellulose-based polymers adsorb enzymes effectively that requires washing them with solutions of high ion strength. This procedure can cause dissociation of polymers forms of enzymes that is why cellulose as a carrier is unsuitable in this case.

✓    The carrier mechanical properties, its mechanical strength and physical form in particular, should be taken into consideration. For example, the carrier has to be capable of forming films if a film of immobilized enzyme is needed.

If all the factors mentioned meet necessary conditions, the immobilization method can be considered as appropriate.

### 1.5. Conductometric enzyme sensors

The first conductometric biosensor for urea determination has been described in [19]. It was a device consisting of a silicium substrate with a pair of gold interdigitated and serpentine electrodes. The experiments were carried out in both a laboratory and clinics; the biosensor response to urea was in the range of $0.1 - 10$ mM in imidasole buffer, pH 7.5. The $K_M$ of immobilized enzyme was higher than for the native one; the authors explained it as a result of diffusion limitation. A comparison of the data obtained by the biosensor in the laboratory with the results of conventional clinical tests showed good agreement (the correlation coefficient was higher than 0.99).

Similar conductometric biosensors have also been used as a multisensor [33]. Urease was immobilized on the surface of the first electrode pair in the gel layer; on the second pair there was L-asparaginase; on the third pair – a three-enzyme system "urease-creatinase-creatininase". This sensor was used for the determination of urea, L-asparagine and creatinine, respectively. The sensor was tested both separately with each of the substrates, and in multi-substance mode the kinetic and calibration curves were determined.

In Table 1.1 the results of development in the field of conductometric enzyme biosensors during the last decades are presented. It is obvious that conductometric transducers have mainly been used for elaborating biosensors for urea determination.

The multisensor described in [43], consisted of a conductometric biosensor for urea analysis combined with an amperometric biosensor for glucose determination. It was highly selective and simple to operate, and was used in clinics.

Mikkelsen *et al.* have characterized conductometric biosensors for urea and D-amino acids determination [35]. The enzymes urease and D-amino acid oxidase were used. The minimum detection limit for urea concentration was 5 µM; the linear dynamic range was of three orders. The dependence of response on buffer capacity was studied. While the sensor for D-amino acid analysis was being developed, the D-amino acid oxidase was co-immobilized with catalase, since hydrogen peroxide, being the product of the enzymatic reaction, and is the inhibitor of D-amino acid oxidase. A comparative analysis of using copper and platinum electrodes, as well as different buffer solutions showed that the platinum electrodes and glycine buffer were preferable. An optimal pH of the sensor for D-amino acids and its selectivity towards various amino acids were determined. The sensor showed stable results during its 33-days operation.

Two types of thick-film conductometric biosensors for urea determination are the subject of [44]. The first type was manufactured by printing two interdigitated electrodes onto an $Al_2O_3$ substrate using platinum paste, while the second, consisting of four silver-palladium electrodes in parallel – by the "green tape" technology. Urease was immobilized by covalent binding in albumin gel. The response time for both biosensors was about 10 min. The dynamic ranges for the first biosensor were: 0.1 – 50 mM urea, the linear part of 0.1 – 4 mM, for the second 10 µM – 5 mM urea, the linear part of 10 – 350 µM. These biosensors were shown to suit medical analysis.

The conductometric biosensor based on inhibition analysis, first described in [34], was intended for the determination of organophosphorous pesticides. As a sensitive element, the enzymes acetyl- and butyrylcholinesterase were used. The sensor sensitivity to different pesticides (diisopropyl fluorophosphate, paraoxon-ethyl, paraoxon-methyl, trichlorfon) was investigated, the minimal detection limits for inhibitor concentrations

*The data on development of different conductometric biosensors*

| № | Analyte | Enzyme | Reference(s) |
|---|---------|--------|--------------|
| 1 | Total protein | Trypsin | [45] |
| 2 | Creatinine | Creatinine deiminase | [33] |
| 3 | L-asparagine | L-asparaginase | [33] |
| 4 | Glucose | Glucose oxidase | [45] |
| 5 | Hydrogen peroxide | Peroxidase | [46] |
| 6 | D-amino acids | D-aminoacidoxidase | [35] |
| 7 | Acetylcholine | Acetylcholinesterase, | [47] |
| 8 | Butyrylcholine | Butyrylcholinesterase | [47] |
| 9 | Penicillin | Penicillinase | [45] |
| 10 | Uric acid | Uricase | [48] |
| 11 | Urea | Urease | [33, 35, 45, 48] |

were the following: $5 \times 10^{-11}$ M for diisopropyl fluorophosphate, $10^{-8}$ M for paraoxon-ethyl, $5 \times 10^{-7}$ M for paraoxon-methyl, and $5 \times 10^{-7}$ M for trichlorfon. The dependence of biosensor response on the duration of the transducer incubation in the pesticide solution was studied. The possibility of enzyme reactivation in the membrane by means of reactivator pyridine-2-aldoxime methiodide was shown. The conclusion was drawn that the biosensors described could be used for the analysis of organophosphorous pesticides in aqueous solutions.

The potential of a conductometric urease biosensor for the determination of heavy metal ions was demonstrated in [2]. The inhibition activities of heavy metals towards urease varied as follows:

$Hg^{2+} > Cu^{2+} > Cd^{2+} > Co^{2+} > Pb^{2+} > Sr^{2+}$; reactivation of the inhibited enzyme with EDTA was shown to be probable.

The conductometric biosensors applied to analyzing total solution toxicity at parathion-methyl photodegradation were presented in [49]. The results obtained were compared with the data from traditional high-sensitive method of HPLC and from the Lumistox® device (Germany) for toxicity determination. The solution toxicity was shown to increase dramatically as pesticide photodegradation began, the toxicity remained once the parathion-methyl degradation had completed. However, the authors do not oppose the biosensor method to others, but consider it as an additional fast way for early screening of numerous samples.

The comparative analysis of the operational characteristics of enzyme biosensors for penicillin determination based on conductometric planar electrodes, on the one hand, and pH-sensitive field-effect transistors, on the other, demonstrated [50] similarity in their analytical parameters: both have short response time and high operational stability. However, planar conductometric electrodes are preferable from the technological aspect, because they are cheaper and easier in manufacture, and, therefore, promising for practical use. The possibility of selecting the necessary dynamic range of the transducer operation by varying medium buffer capacity was also shown.

Therefore, the application of conductometric measuring method to continuous recording in the course of enzyme processes is thoroughly examined and analyzed regarding both standard conductometers and conductometric enzyme biosensors. As compared with conventional methods of biochemical analysis, the method considered is universal, features higher accuracy and low cost.

The conductometric biosensors also have some advantages over other types of transducers. First, they can be produced through inexpensive thin-film standard technology. This, along with using an optimized method of immobilization of biological material, results in considerable decrease in both primary cost of devices and the total price of analyses. For integral microbiosensors it is easy to perform a differential measurement mode, thus compensating for external effects and considerably increasing measurement accuracy.

The data is convincing evidence of great potential of conductometric biosensors. However, it is still rather a novel trend in the field of biosensors, which is why the development of commercial devices has a promising future.

### 1.6. Application of biosensors for environmental monitoring

Monitoring of the environment for the presence of compounds which may adversely affect human health and local ecosystems is a fundamental part of the regulation, enforcement and remediation processes which will be required to maintain a habitable environment. Because of the diversity of current environmental monitoring methods, the choice of techniques used for sample collection, extraction and analysis depends on the compound or compounds of interest and the matrix they contaminate. Although classical analytical techniques are being continually refined and improved, these laboratory-based methods are, for the most part, relatively expensive and time-consuming. So biosensor technology lends itself to fast, economical and portable analysis, these devices may provide solutions for some of the problems currently encountered in the measurement of environmental contaminants.

More than 1000 chemical compounds have been identified at hazardous waste sites alone and there may be thousands of unidentified pollutants. The majority of compounds found on the Agency for Toxic Substances and Disease Registry's (ATSDR's) priority list are volatile organic compounds; however, other prominent chemical classes include inorganic, phenols, phenoxy acids, polyaromatic hydrocarbons, nitrosamines and aromatic amines [51]. In many cases, identification and quantitation of specific compounds in mixtures of these chemicals must be made in media such as air, water, soil and sludge as well as biological matrices such as blood, saliva and urine. Although most biosensors must operate in aqueous environments, the development of gas-permeable membranes and appropriate soil extraction procedures can extend the range of media accessible to analysis by these devices.

Biosensors research over the last decade has been focused primarily on clinical applications. Only recently, driven by the need for better methods for environmental surveillance, reports concerning biosensors for these applications become more prevalent in the literature. Biosensors reported for environmental applications measure a fairly broad range of compound classes (Table 1.2)

For environmental monitoring, there are several general areas in which biosensors may have distinct advantages over current analytical methods. Miniaturization of biosensors has, in the clinical field, led to the development of economical (i.e. disposable or reusable) sensor modules which can be detached from the display module. For example, a blood glucose analysis can typically be made for less than 1 EUR. This is an attractive feature for environmental monitoring, considering that the average cost of a laboratory analysis for an environmental sample can range from 100 to 200 euros. The development of portable biosensors could also eliminate the time, expense and chain-of-custody considerations required to transport samples to a laboratory. Other possible advantages of biosensor-based assays include the measurements of analytes in complex matrices with a minimum of sample preparation, operation of the biosensor by untrained personnel, completion of the analysis in less than 1 hour and continuous or remote monitoring capabilities.

There is a critical and growing need for more cost-effective techniques for identification and quantitation of pollutants in complex environmental matrices. For biosensors to become competitive their characteristics must be exploited to fill existing technology gaps left by classical laboratory analysis and emerging field screening methods such as portable gas chromatography, mass spectrometry, chemical sensors, ion mobility spectrometry and immunoassay test kits.

This is not a trivial problem considering that out of hundreds of existing designs, relatively few biosensors have been developed into commercial products. As chromatographic, immunoassay and spectroscopic-based field screening assays improve; biosensors have to meet or exceed the quality of data produced by these methods for the specific analyte of interest. Issues which must be addressed in the development of biosensors include detection limits, dynamic range, positive and negative controls, interferences and matrix effects.

As problematic toxic waste repositories are identified and remediation procedures are implemented, "real-time" monitoring assays become crucial for directing appropriate clean-up strategies. Existing and emerging chromatographic and spectroscopic field screening technologies appear well suited to the detection of certain pollutants such as volatile organic compounds and heavy metals. Furthermore, immunoassays have been demonstrated for a wide range of insecticides (e.g. chlorinated hydrocarbons and organophosphates), herbicides (e.g. thiocarbamates, phenoxyaliphatic acids, bipyridiliums and triazines), fungicides (e.g. benzimidazoles, acylalanines and triazoles) and rodenticides (e.g. coumarins). Biosensors can exploit the sensitivity and specificity afforded by antibodies used in these assays and take advantage of technological design improvements in areas of speed, simplicity and 'user friendliness'.

Another area in which biosensors may find a niche involves continuous and/or remote monitoring of industrial or agricultural groundwater or runoff. Chemical sensor technologies have made recent progress in this area. These devices use mass, optical, electrochemical or thermal sensors coupled to chemically selective coatings. Chemical sensors and biosensors share many of the same characteristics (e.g. size, speed and cost of analysis). Continuous and/or remote monitoring applications, however, have specific requirements in that unattended operation over extended periods is needed. Chemical sensors have the advantages of ruggedness and durability for this application; however, biosensors have the advantages of sensitivity and selectivity. Challenges for biosensors in this area include the stabilization of the biologically active membranes and coatings and the development of methods for continuous delivery of the biological macromolecules into the system.

Biosensors have attracted tremendous interest and attention in science since the late 1970s because they have a wide range of possible applications. After more than a decade of intense research, however, most effort still remains at the level of research curiosities in universities and research institutes. At present, there are only a few biosensor technologies, among hundreds of existing designs that have been developed into commercial products. None of these biosensors have been commercialized for environmental applications. Although each sensor has its own performance requirements and design specifications, there are several common obstacles to commercialization of biosensors in general.

First, the performance characteristics such as sensitivity, dynamic range, reproducibility, and stability, for sensors demonstrated in the research stage have not quite met the needs of users. Furthermore, the performance of the biosensor is further reduced in the development and manufacturing stages as a result of mass production. Therefore, to make a practical device, it is important to keep chemical and physical design characteristics simple, using proven technologies wherever possible. The second obstacle is the lack of effort toward system integration and user-interfacing. The sensor concepts need to be taken one step further by building a prototype model that can be easily operated by untrained users. The third obstacle involves the competitiveness of the biosensors. Commercialization becomes difficult or impossible if no definite advantages over other competing technologies are achieved. Finally, probably the most important factor responsible for the slow commercialization of biosensors is the market size. Because of the complexity of these devices, millions of euros must be invested to develop a biosensor.

Therefore, until profitable marketplaces are realized, private companies will not be motivated to investigate and transfer the biosensor technologies to products. Issues related to competitiveness and market size for most biosensor technologies are still being defined by current research and market developments.

It seems that biosensors will have a more important role in various sectors of the analytical industry, such as environmental monitoring.

## 1.7. Conclusions

The fundamentals of conductometric sensors have been considered, taking into account their potential application. A number of peculiarities of the most advanced microtechnologies used in production of conductometric biosensors have been presented, experimental measurement circuits demonstrated. The achievements in the field of development of enzyme conductometric biosensors have been described in detail, the prospect of their application evaluated. Some environmental monitoring systems based on biosensors have been exposed along with their practice.

There currently exists a clear and increasing need for environmental field analytical methods which are fast, portable, and cost-effective. Biosensors with their versatility may find a number of specific applications among the variety of other methods and technologies currently competing for this expanding market. In fact, a variety of enzyme, antibody, and microbial-based biosensors using optical, electrochemical, and acoustic-signal transducers have been reported to measure a significant number of environmental pollutants from a variety of compound classes.

Nevertheless, there are a considerable number of technical challenges and practical obstacles which must be met before any of these field analytical methods find widespread acceptance and use. During the development of biosensors for environmental monitoring applications, it is important that there exists a free flow of creative ideas concerning the use of diverse biological recognition elements, alternative operating formats, and innovative signal transducers. Nevertheless, it is imperative that sufficient thought also be given to issues such as the requirements of potential environmental applications, the ability easily to adapt and manufacture laboratory prototypes, and the presence of a sufficient market to warrant the investment required for commercialization of these devices. Although developing biosensors for environmental applications is not a trivial task, there appears to be sufficient evidence that biosensors can be configured to be selective, sensitive, and inexpensive to manufacture. It is further likely that with appropriate development, testing and commercialization, biosensors can have a significant impact on reducing costs and increasing the efficiency of certain environmental monitoring applications.

**Table 1.2**

*The data on development of different biosensors for ecological monitoring*

| № | Analyte | Biosensitive element | Reference(s) |
|---|---------|---------------------|--------------|
| 1 | Organo-phosphorous pesticides | Acetylcholinesterase, Butyrylcholinesterase | [52] |
| 2 | Heavy metal ions | Urease | [2] |
| 3 | Formaldehyde | Alcoholoxidase | [45, 53] |
| 4 | 4-Chlorphenol, | acetylcholinesterase | [52] |
| 5 | Phenolics compaunds | Tyrosinase | [54] |
| 6 | Ammonia | Nitrifying bacteria | [55] |
| 7 | Polycycl. Arom. Hydrocarb. | Antibodies | [56] |
| 8 | Nitrates | Nitrate reductase | [57] |
| 9 | Metals | Mercuric reductase | [58] |
| 10 | Herbicides | Antibodies | [59] |
| 11 | Insecticides | Acetylcholinesterase | [60] |
| 12 | Biochem. O Demand (BOD) | *Bacillus subtilis* | [61] |
| 13 | Bacterial contamination | Amperometric mediators | [62] |

# Chapter 2

### Chapter 2. Technological and methodological fundamentals of planar conductometric transducers

#### Résume en français du chapitre 2

L'objectif de ce deuxième chapitre est d'analyser le fonctionnement des transducteurs conductométriques et d'analyser les caractéristiques des microélectrodes en fonction des matériaux, de leur dimensions et de leur géométrie.

Comme nous l'avons souligné dans le premier chapitre, le biocapteur se' constitue de deux parties clés : (i) un élément biosélectif, qui est responsable de la reconnaissance sensible suffisante et de la traduction de l'information du domaine biologique dans un signal sortant chimique ou physique, et (ii) un transducteur électrochimique, responsable pour la traduction du signal dans un domaine électrique et la transformation dans l'information accessible analytiquement. Le point le plus important dans l'élaboration des biocapteurs, à côté du choix optimal du transducteur, est la sélection du design, des dimensions et du matériel de transducteur la plus efficace. Par conséquent, nous avons choisi cet aspect en tant que l'objet de notre investigation dans ce chapitre.

Une électrode interdigitee à film mince est le plus efficace pour le développement des biocapteurs conductimétriques. Le transducteur conductométrique à film mince est un transducteur miniature qui contient deux électrodes interdigitales à film mince, conçut pour mesurer la conductivité de la couche de surface des électrodes. Quand le transducteur est couvert avec une membrane biologique active, il se transforme en un biocapteur conductométrique.

Nous avons proposé le rang prioritaire de différents matériaux d'électrodes sensibles par rapport à leur utilisation dans les biocapteurs conductométriques. Il est le suivant:

platine>or>nickel>cuivre>chrome>titane>aluminium

Nous avons démontré que le matériel de la substance non-conductive n'a pas d'effet sur la sensibilité du transducteur. Par conséquent, il peut être choisi en fonction de la facilité de sa production et de son prix. Les dimensions caractéristiques des parties d'une électrode sensible ne sont pas le paramètre déterminant. C'est pourquoi la technique de lithographie régulière peut être utilisée pour déposer les électrodes dans les biocapteurs conductometriques.

Les résultats obtenus dans la présente recherche ont été utilisés dans un nombre d'organismes de production des transducteurs conductométriques. Les transducteurs fabriqués à Kyiv Radio-usine (Ukraine) et Institut de Chémo- and Biocapteurs (Muenster, Germany) ont démontré les meilleures caractéristiques et ont été recommandés pour le développement des biocapteurs conductométriques. Nous avons proposé les dispositifs expérimentaux pour l'investigation de biocapteurs conductométriques ainsi que l'appareil portable développé sur sa base.

En général, les résultats de cette étape d'investigation se résument à des points suivants :

(a) nous avons soigneusement analysé les caractéristiques analytiques et l'efficacité opérationnelle de différents transducteurs électrochimiques ;

(b) le design bénéfique, les matériaux optimaux et le dispositif de mesure ont été recommandé pour chaque type de transducteur. Ces recommandations ont été appliquées en Ukraine et à l'étranger (pays Européens) chez les fabricants des transducteurs ;

(c) nous avons constaté que les transducteurs conductométriques sont les plus simples du point de vue technologique. Par conséquent, ils sont avantageux en terme de leur production et système de mesure.

## 2.1. Introduction

A part of this chapter was published in [63].

As it has been stated in the previous section, any biosensor consists of two core units, (i) **a bioselective element,** responsible for sufficiently sensitive recognition and translation of the information from a biological domain into the chemical or physical output signal, and (ii) **an electrochemical transducer** responsible for the signal translation to an electric domain and transformation into the analytically available information.

The most important issue in elaborating biosensors, is selection of the most effective design, dimensions and material of the transducer. It became a subject of priority in further overall investigation.

An interdigitated thin-film electrode is the most effective for development of conductometric biosensors. A thin-film conductometric transducer is a miniature transducer consisting of two interdigitated metal electrodes, assigned to measure conductivity of the solution layer at the surface of electrodes depending on their characteristic dimensions. Being covered with a biologically active membrane, the transducer turns into a conductometric biosensor [64].

## 2.2. Simulation by equivalent electronic circuits

In general, a thin-film conductometric biosensor can be schematically presented as in Fig. 2.1.

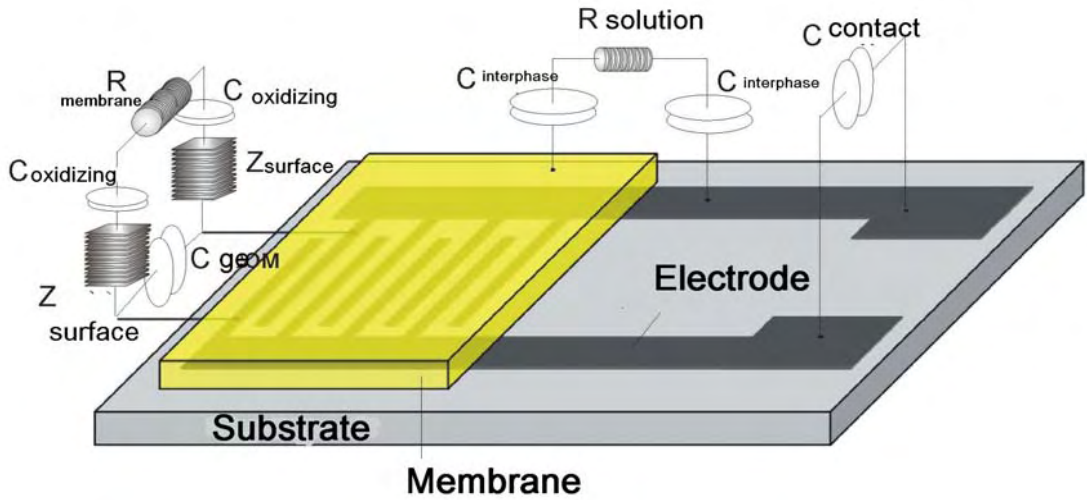

**Fig.2.1.** *Scheme of interdigitated thin-film planar conductometric biosensor [10].*

Physical and chemical processes in an electrochemical cell with conductometric biosensor can be simulated according to the equivalent scheme (Fig. 2.2).

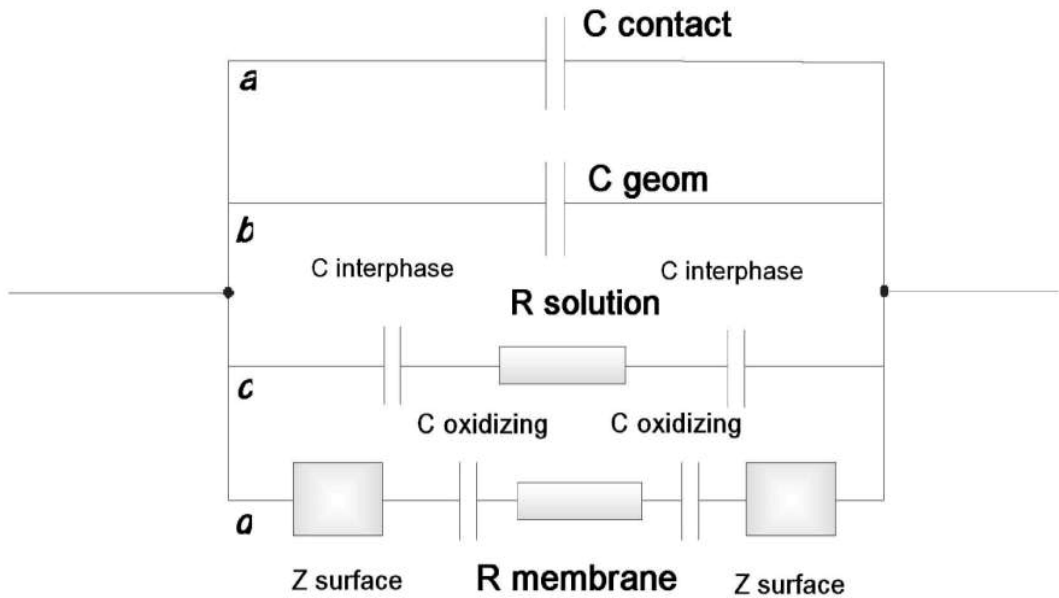

**Fig. 2.2.** *Equivalent scheme simulating physical and chemical processes in an electrochemical cell with conductometric biosensor [10].*

Let us examine the scheme:

The circuit $a$ – $C_{contact}$ is a capacity of contact planes and wires which connect electrodes to the measurement system. It depends generally on the material of planes and wires and varies from 0.1 to 1 pF, thus, this circuit can be neglected.

The circuit $b$ – $C_{geom}$ is a geometrical capacity of metal sensitive parts of the transducer. Depending generally on their material, geometry and dimensions it changes from 0.1 to 1 pF and can be neglected as well.

The circuit $c$ is the system of electrochemical impedance describing the out-of-membrane region. It consists of two equal capacities $C_{interphase}$ responsible for the membrane-solution interphase, and the resistance, $R_{solution}$, simulating solution conductivity of this region.

The circuit $d$ is the system of electrochemical impedance describing the inside-membrane region. It consists of two equal surface impedances capacities $Z_{surface}$, simulating the membrane-solution interphase, two equal capacities $C_{oxidizing}$ simulating the electrode material oxide, and resistance $R_M$, simulating conductivity of the inside-membrane region.

In our case, the solution swelled membrane is 200 µm thick, while the size of active electric region of electrodes equals three electrode characteristic dimensions, *i.e.* less than 210 µm as the maximal characteristic dimension of the electrodes used is 70 µm. This means that the circuit $c$ is negligible as well. Even more this consideration concerns experiments with conductometric transducers without membrane.

Thus, rather a complicated equivalent scheme simulating the processes in the cell with conductometric transducer turns in our case into quite simple one and consists of the circuit *d* only. It means that the sensor response is mainly determined by electrochemical impedance of the "metal electrode-solution" system (Fig. 2.3).

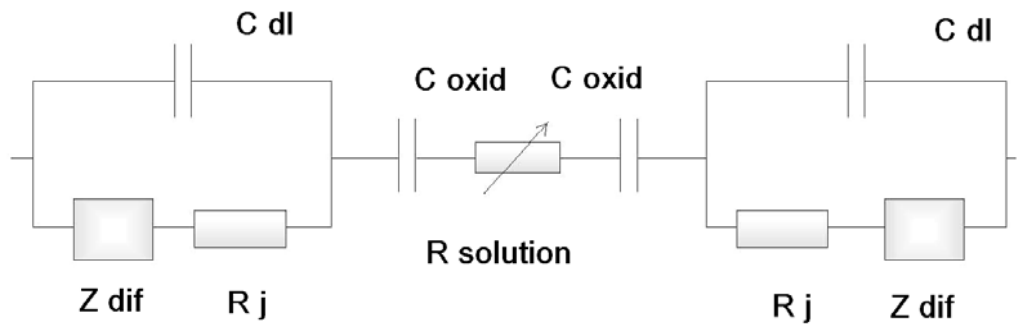

**Fig. 2.3.** *Simplified equivalent electric circuit simulating processes in a cell with conductometric transducer [10]*

Thus, the electrochemical impedance of the conductometric transducer-solution system consists of the following:

1. $C_{dl}$ – capacity of double electric layer which actually equals 10 – 20 μF/cm$^2$, determines quantitatively the oriented dipoles at the electrode surface and does not depend on current frequency;

2. $R_j$ – junction resistance which is responsible for chemical polarization due to electrochemical reactions at the electrode surface and is frequency-independent;

3. $Z_{dif}$ – diffusion, or Warburg, impedance which is responsible for concentration polarization due to the ion diffusion from interface into electrolyte volume, and is frequency-dependent;

4. $C_{oxid}$ – capacities of oxide of electrode material;

5. $R_{solution}$ – resistance of electrolyte which simulates the resistance of solution inside membrane.

If frequency is higher than 10 Hz, the diffusion impedance is negligible while the solution resistance, capacity of double layer, and junction resistance make an essential contribution into the system general electrochemical impedance [64].

### 2.3. Design of thin-film interdigitated electrodes

The following transducers were used and investigated:

1. Conductometric transducers, manufactured with our laboratory templates at R&D Microdevice Institute (Kyiv, Ukraine), using integral planar technology. The sensor was of the following design: substrate (30 x 5 mm) made of pyroceramics (Al$_2$O$_3$) or silicon with isolating layer of silicon nitride, with two equal pairs of gold interdigitated electrodes, obtained by vacuum deposition. In order to achieve better adhesion, the titanium sublayer (0.1 µm) was used. The distance between digits was 40, 20 and 10 µm.

2. Transducers, produced according to our recommendations at Kyiv Radioplant (Ukraine). Sensor included polyceramic substrate (40 x 5 mm), with gold, nickel, copper, and chrome electrodes applied photolithographycally or by vacuum deposition. As sublayer for improved adhesion chrome layer of 0.1 µm thick was used. The distance between digits was 70 µm.

3. Transducers developed at the Institute of Microelectronics, Newchatel, Switzerland. Silicon oxide substrate (3 x 10 mm) with gold electrodes applied photolithographycally or by vacuum deposition was used. The width of digits and the distance between them was 0.5, 2, 5, and 10 µm. The chip was glued to the base of 5 x 100 x 1 mm.

4. Sensor, produced according to our recommendations at the Institute of Chemo- and Biosensorics, Munster, Germany. 40 x 5 mm glass substrate with platinum electrodes was used. The width of digits and the distance between them was 10 µm.

Fig. 2.4 presents the general view of the mentioned transducers.

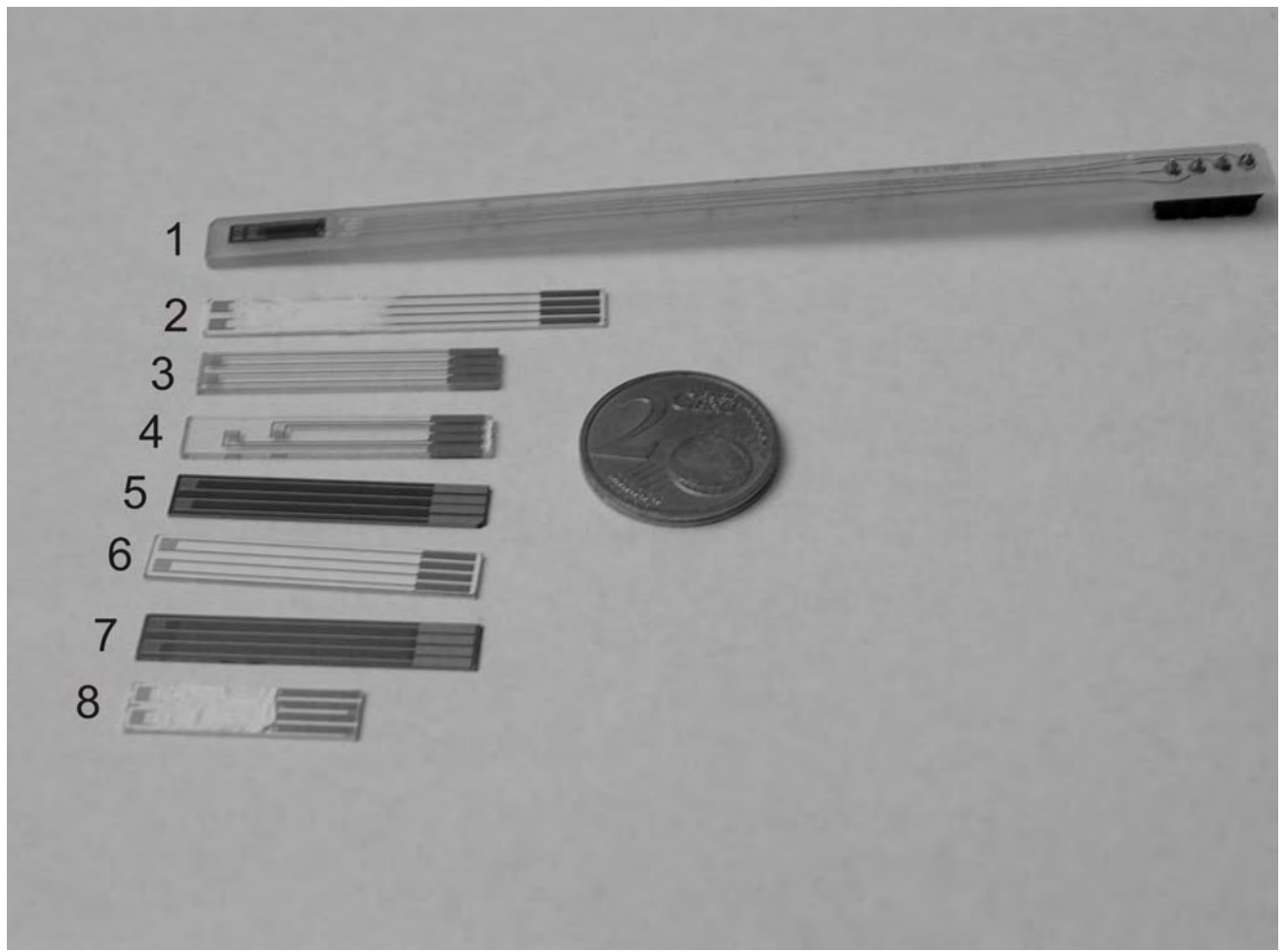

**Fig. 2.4.** *General view of conductometric transducers produced in R&D Microdevice Institute, Kyiv, Ukraine (5, 6, 7), Kyiv Radioplant, Ukraine (2, 8), Institute of Microelectronics, Newchatel, Switzerland (1), Institute of Chemo- and Biosensorics, Muenster, Germany (3, 4)* [63]

Middle parts of electrodes were protected by insulating layer (polyamide, photoresist, carbon oxide or epoxy resin). Each pair of electrodes served as a conductometric transducer.

### 2.4. Experimental device and methods for impedance measurement

Investigation of characteristics of thin-film conductometric transducers as components of the biosensors using impedance (admittance) spectroscopy was carried out to determine the options of their effective application. The reactive and active parts of impedance (admittance) vs. frequency were plotted.

The impedance measurements were performed by the device (Fig. 2.5) on the basis of the impedance-spectrometer manufactured by *Schumberger Co.*, Germany [65].

The signal from impedance-spectrometer was logarithmically varied from 0.01 Hz up to 1 MHz according to the problem posed. The impedance values were registered by PC, processed and displayed as impedance or admittance curves, and recorded as data files.

All measurements were performed at room temperature in various aqueous solutions, namely, distilled water and potassium-phosphate buffer solution ($KH_2PO_4$-NaOH, pH 7.4). The electrolyte conductivity was changed, if needed, by addition of various aliquots of 4 M KCl.

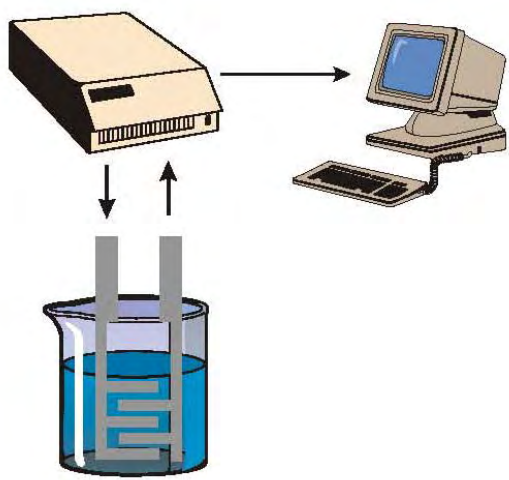

**Fig. 2.5.** *Scheme of device for impedance measurement using impedance-spectrometer manufactured by Schumberger Co., Germany [10]*

**2.5. Investigation of transducer characteristics depending on electrodes material, dimensions and geometry**

### 2.5.1. Material of interdigitated electrodes

Material of interdigitated electrodes should be optimized towards its price and sensitivity to varying conductivity of the solution tested. The characteristics of interdigitated electrodes of different materials were compared by impedance spectroscopy. Series of structures of gold, nickel, chrome, copper, platinum, titanium, and aluminum were examined. The dependence of real part of admittance of the electrochemical cell on frequency presented in Fig 2.6 was measured in distilled water and 1 mM KCl solution for electrodes of the materials mentioned with the same characteristic dimensions. At increasing medium conductivity, the system total admittance and, correspondingly, its real component, conductivity increases. As it can be seen, all admittance curves, except of copper case, have the resembling behavior. Copper electrodes show oxidation process during the experiment.

Further, the electrodes sensitivity to the changes in the solution conductivity (KCl concentration) was examined, the conductivity changes during the enzyme reaction being simulated. The results presented in Fig. 2.7. concern the dependence of the admittance active component of the transducer on the basis of platinum electrodes on frequency, and testify evidently to the conductivity changes at high frequencies.

The same form of curves obtained in similar experiments for all the transducers tested, confirms the accuracy of the equivalent circuit proposed. The data obtained for 100 kHz were used to plot the response dependence on KCl concentration in distilled water for conductometric transducer with electrodes made of different metals (Fig. 2.8).

Evidently, the platinum electrodes are the most sensitive to the changes in KCl concentration. The sensitivities of gold, copper, and nickel electrodes are rather less and alike for the whole group while the chrome and titanium ones, and especially aluminum electrodes demonstrate considerably lower (one order) sensitivity. The KCl concentration of 5 – 10 mM is sufficient for them to be entirely saturated towards their conductivity. It makes them unusable as conductometric transducers in enzyme biosensors, since the latter should be utilized in biological liquids, which are in fact featured in high background conductivity.

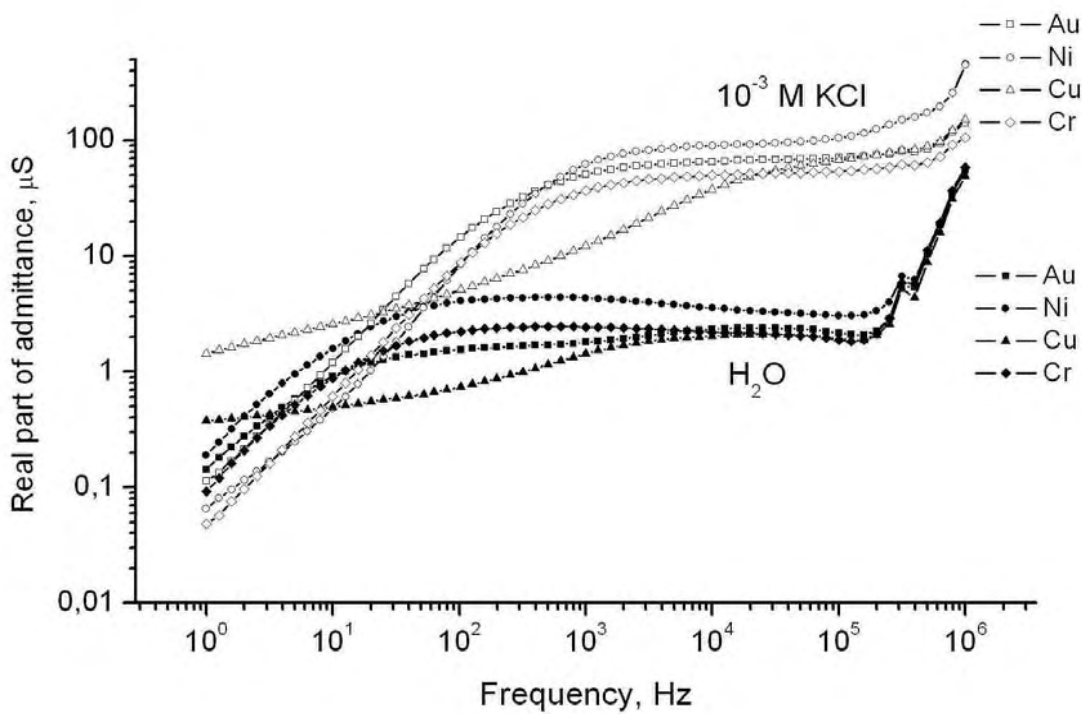

**Fig. 2.6.** *Dependence of admittance real part on current frequency for electrodes of different materials. Electrodes distinctive size was 10 micrones characteristic dimension. Electrode substrate was silicon oxide. Signal amplitude was 10 mV. Measurements were conducted in bidistilled water and in 1mM KCl solution in bidistilled water.*

At the same time, such base cheap metals, as nickel and copper are quite suitable for conductometric enzyme biosensors, since the analytical characteristics of transducers on their basis are similar to those based on noble metals. However, copper electrodes are negatively featured towards time factor because of fast oxidation of their surface and, thus, reducing their sensitivity. Described, in figures 2.6 and 2.8, experiments were realized with new copper electrodes without oxide film. After several measurements they obtain oxide film and have the behavior similar with a not noble metal based electrode.

Therefore, the line of materials with regard to priority usage in conductometric biosensors is as follows:

*platinum>gold>nickel>copper>chrome>titanium>aluminum*

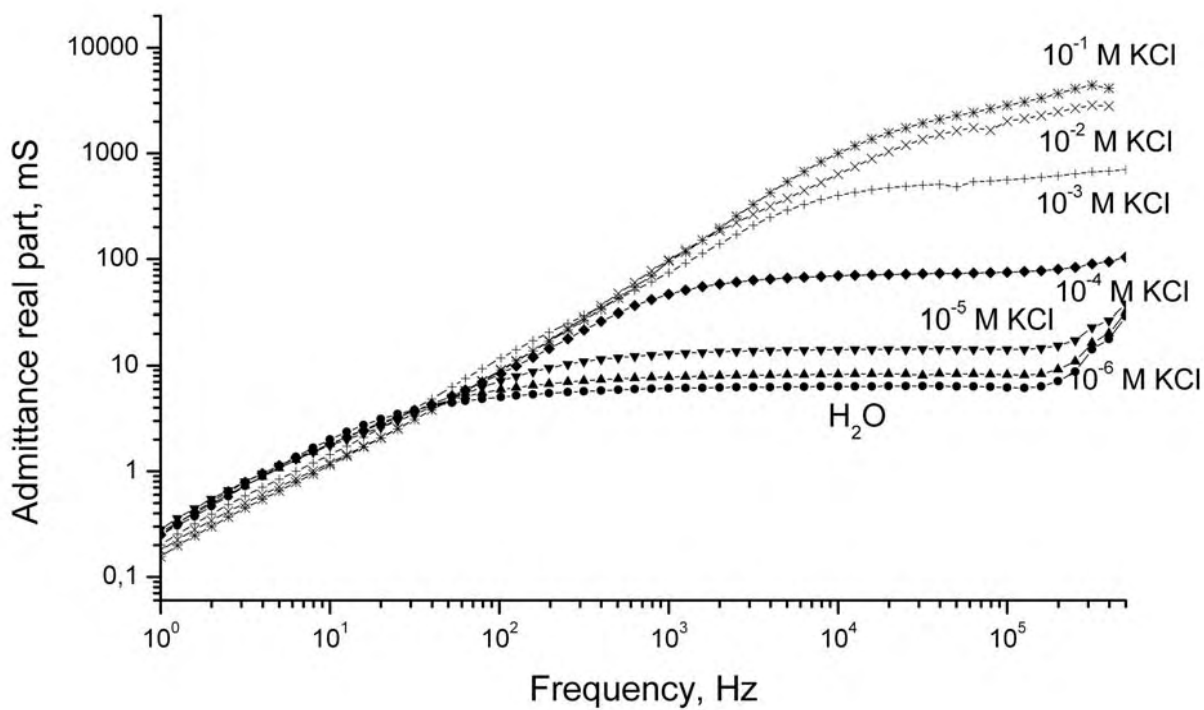

**Fig. 2.7.** *Dependence of the admittance real part on current frequency for different KCl concentrations. Measurments were perfomed with 10 microns characteristic dimension platinum conductometric transducer. Electrode substrate was silicon oxide. Signal amplitude was 10 mV. Measurments were conducted in bidistilled water*

### 2.5.2. Material of transducer substrate

The transducers with gold electrodes deposited on glass, pyroceramic and silicon oxide substrates were studied. The results of investigations of solutions conductivity with various ion strength by transducers with different substrates are presented in Fig. 2.9.

It is apparent that substrate material has no effect on the transducer sensitivity to changes in the medium conductivity. The only requirement to the substrate material is its non-conductivity. From technological point of view, glass is preferable since it is the cheapest and the least fragile material.

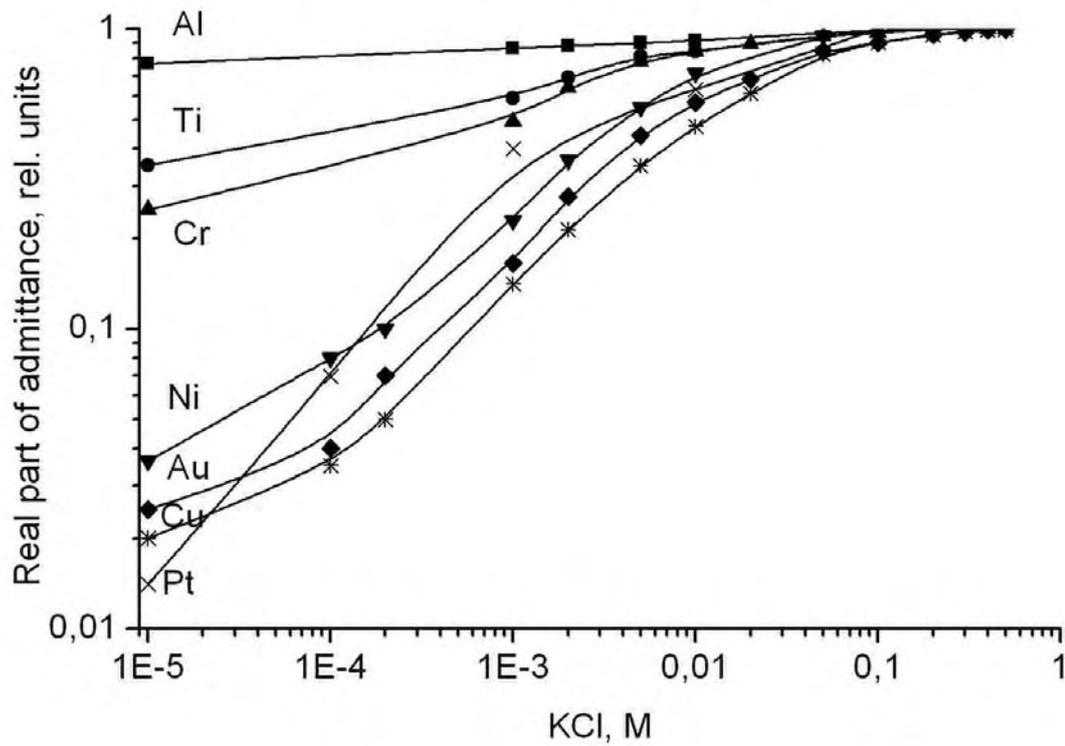

**Fig. 2.8.** *Dependence of sensor response on KCl concentration in distilled water for 10 microns characteristic dimension electrodes made of different metals: aluminum, titanium, chrome, nickel, gold, copper, platinum. Electrode substrate was silicon oxide. Signal frequency and amplitude were 100 kHz and 10 mV respectively. Measurements were conducted in bidistilled water*

### 2.5.3. Electrode characteristic dimensions

The characteristic dimensions were considered in the experiments on effect of dimensions of interdigitated electrodes digits and the distance between them. Two groups of transducer were tested: the first consisted of electrodes with characteristic dimensions of 20, 40, and 70 µm and active area of 2.25 mm$^2$, the second, 10, 5, 2, and 0.5 µm and active area of 1.5 mm$^2$, respectively.

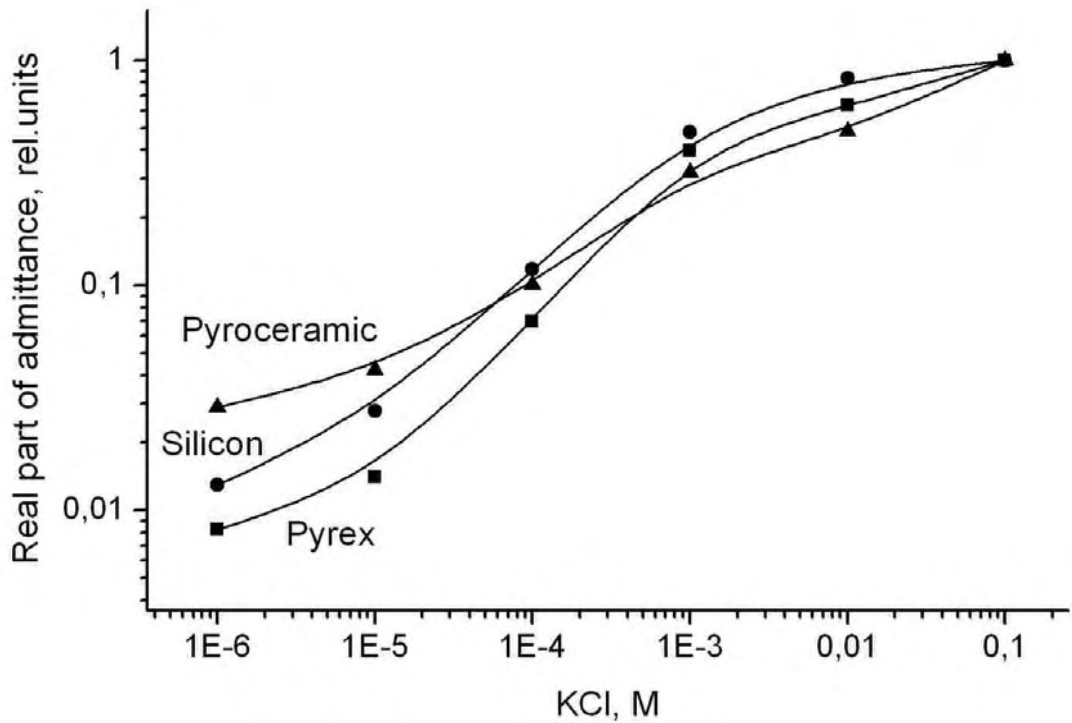

**Fig. 2.9.** *Dependence of sensor response on KCl concentration for 10 microns platinum electrodes deposited on various substrates (1 – pyroceramics, 2 – silicon oxide, 3 – pyrex glass). Signal frequency and amplitude were 100 kHz and 10 mV respectively. Measurements were conducted in bidistilled water.*

The sensitivities of transducers for electrodes with various characteristic dimensions are presented in relative units in Fig. 2.10. The sensitivity is seen to trend towards decrease with reducing dimensions of interdigitated electrode digits which agrees with the data obtained by other authors [66, 67]. Thus, a conclusion can be made that, in contrast to earlier consideration, there is no necessity in technologically difficult increasing of digit number by means of reduction of their dimension. The electrodes should be miniaturized by uniform reduction in both electrode working surface and electrode characteristic dimensions. On the other hand, big digits are not recommended (despite higher sensitivity of such transducers) due to increasing influence of membrane thickness. The correlation between membrane thickness, electrode characteristic dimensions, and their active area is the key option at determination of the sensor size. More detailed argumentation on measurement conditions and electrode materials selection can be found in [10].

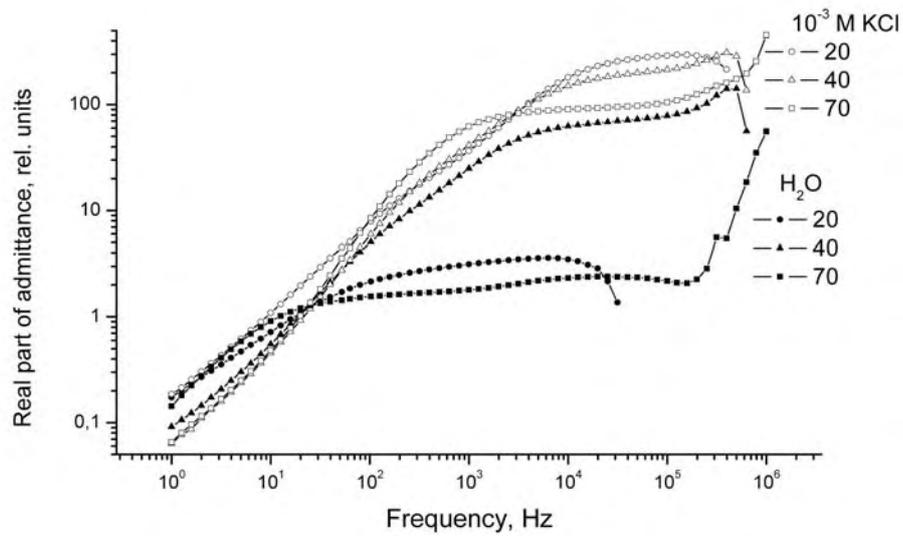

(*a*)

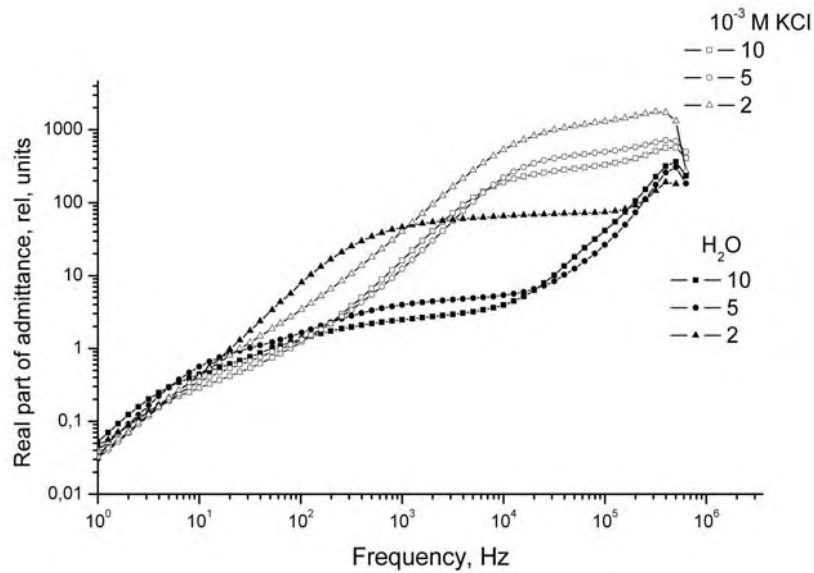

(*b*)

**Fig. 2.10**. *Dependence of sensor response on current frequency for platinum electrodes deposited on silicon oxide substrate with various characteristic dimensions: a – the first group, electrodes of 20 (1), 40 (2), and 70 (3) μm; b – the second group, electrodes of, 2 (4), 5 (5), and 10 μm (6). Signal amplitude was 10 mV respectively. Measurements were conducted in bidistilled water and in 1mM KCl solution in bidistilled water*

### 2.6. Experimental device for measurement by biosensors based on conductometric interdigitated planar electrodes

A scheme of stationary experimental device is shown in Fig. 2.11.

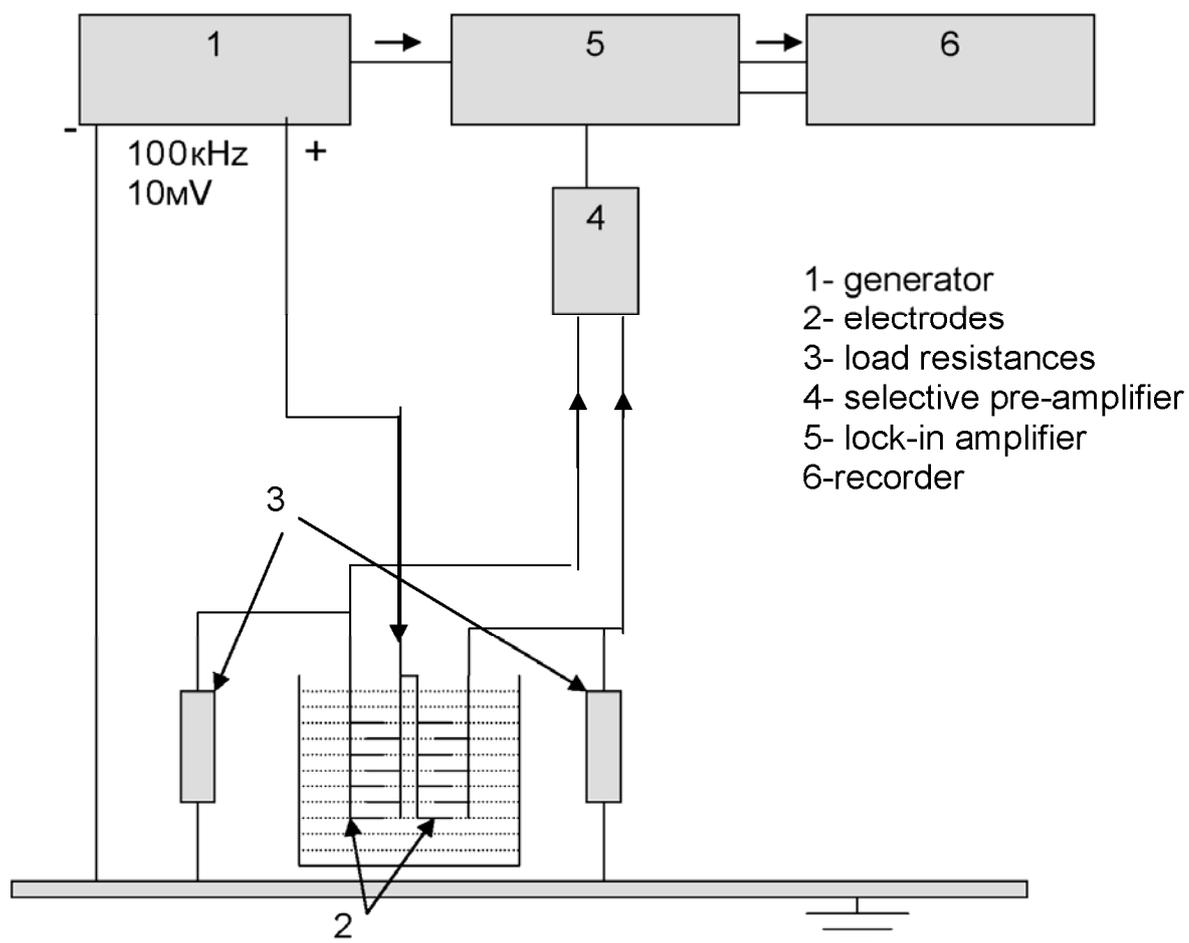

**Fig. 2.11.** *Scheme of experimental device for research of conductometric enzyme biosensors [68]*

An alternating voltage (f = 100 kHz, amplitude 10 mV) was applied from the low-frequency signal generator G3–112.1 (Russia) onto two pairs of interdigitated electrodes placed in the cell with tested solution. An enzyme membrane can be deposited at one electrode pair, while a referent protein membrane – at another one. The circuit load resistance was $R_L$ = 1 kΩ. The electrode output signals entered the differential preamplifier Unipan-233-6 (Poland), where from the differential signal entered the selective nanovoltmeter Unipan-233 (Poland). The necessary component, conductivity, chosen by the nanovoltmeter from the total signal, entered a self-registration apparatus or a personal computer.

A portable device for measurements by the conductometric enzyme biosensor was also designed and made in our laboratory (Fig. 2.12).

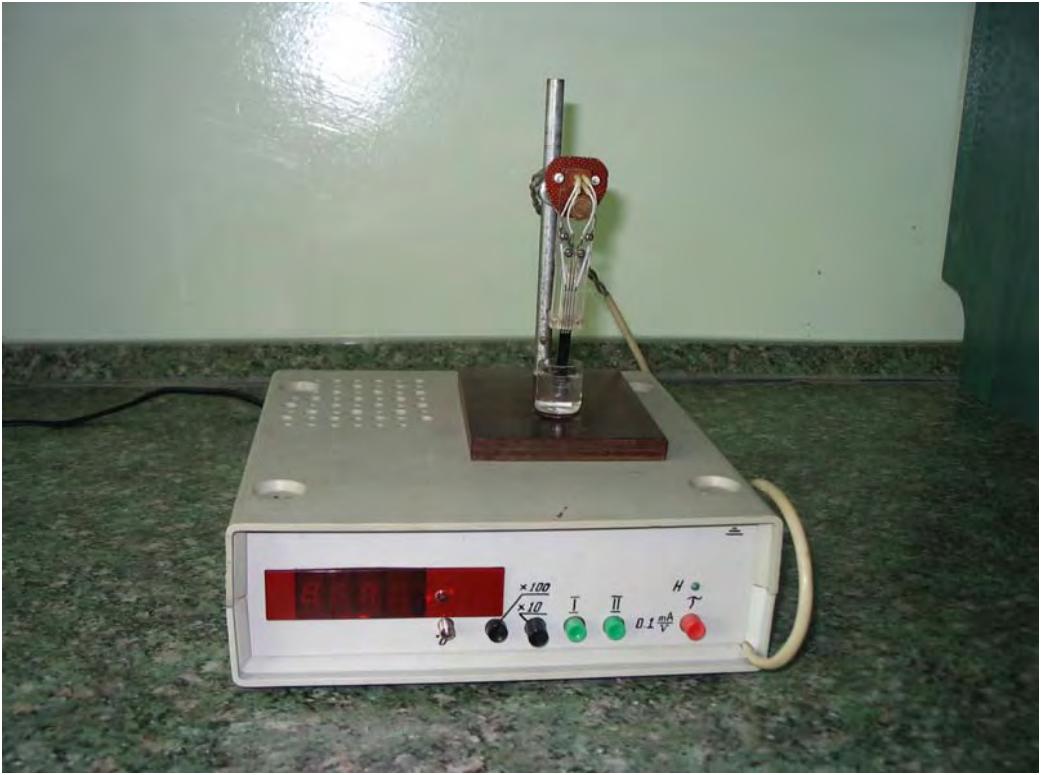

**_Fig. 2.12_** _General view of a portable device for studying biosensors based on_ _conductometric transducers_

The device mentioned can operate as a part of automated measurement equipment and can be considered as a commercial prototype. The generator of sinusoidal vibrations supplies the frequency of 10 kHz and sustains the signal amplitude of 10 mV which allows avoiding the Faraday processes at the electrode surface. The differential operation mode is necessary to increase sensor sensitivity and minimize noise that is why there are two independent input channels of current-voltage conversion and previous amplification in the device. After signal comparison in the differential amplifier, the carrier frequency is separated in the active bandwidth filter, and the signal is detected. The amplifier allows adjusting a necessary zero level for easy cooperation with the external registration apparatus and interior analog-digital converter used for indication of applied voltage. The anticipated measurement sensitivity values are 10 µA/V, 100 µA/V amplified by coefficient 10 or 100. The indicator registration range is 0...12.99.

## 2.7. Conclusion

The priority line of various materials of the electrode sensitive part with regard to their usage in conductometric biosensors is proposed and is as follows:

*platinum>gold>nickel>copper>chrome>titanium>aluminum*

Cheap metals, as nickel and copper are quite suitable for conductometric enzyme biosensors, since the analytical characteristics of transducers on their basis are similar to those based on noble metals. However, copper electrodes are negatively featured towards time factor because of fast oxidation of their surface and, thus, reducing their sensitivity.

The material of non-conducting substance is shown to be of no effect on the transducer sensibility, therefore, it can be chosen by its manufacturability and price. The characteristic dimensions of electrode sensitive parts are not a determinative parameter, and thus, a regular lithography technique can be utilized for electrode deposition in conductometric biosensors.

The results obtained were used by a number of institutions manufacturing conductometric transducers. The transducers produced at Kyiv Radioplant (Ukraine) and Institute of Chemo- and Biosensorics (Muenster, Germany) demonstrated the best characteristics and were recommended for development of conductometric biosensors.

The experimental devices for investigation of conductometric biosensors, as well as a portable measurement apparatus on its basis, were designed.

# Chapter 3

# Chapter 3. Bovin intestinal alkaline phosphatase conductometric biosensor for heavy-metal ions determination

### Résumé en français du chapitre 3

Les activités de recherche sur le développement de biocapteurs susceptibles de détecter la présence de composés organiques ou inorganiques sont à ce jour en plein essor. En effet, ces capteurs offrent de nombreux avantages par rapport aux systèmes analytiques classiques (HPLC/MS, GC/MS, ICP/MS etc) comme par exemple la rapidité de l'analyse, leur simplicité leur sélectivité ou encore leur plus faible coût. Par ailleurs, ils peuvent être utilisés pour réaliser des analyses environnementales aussi bien de l'air que de l'eau, dans le secteur de l'agro-alimentaire, dans le médical ou encore dans l'industrie. En ce qui concerne le secteur environnemental, la détection des métaux lourds représente à ce jour un défi majeur, compte tenu de leur forte toxicité. Il est généralement admis que leur toxicité est lié au fait qu'ils peuvent se fixer sur les groupes thiol de certaines enzymes [69].

A ce jour, différents types de biocapteurs électrochimiques, basée sur l'inhibition d'enzymes par les métaux lourds, ont été proposés dans la littérature. Dans ce chapitre, une synthèse de ces résultats a été réalisée. Parmi ces études seul un nombre limité porte sur l'utilisation de l'alcaline phosphatases (AP) comme enzyme. Pourtant il est connu que l'AP possède une bonne sensibilité vis-à-vis des métaux lourds, et une forte stabilité. Par ailleurs, différentes origines d'AP sont disponibles dans le commerce.

C'est donc dans ce contexte que nous avons essayé de déposer cette enzyme (d'origine bovine dans cette étude) sur les électrodes conductométriques à films minces présentées dans les chapitres précédents et ce afin de mettre au point un biocapteur simple, capable de détecter la présence des ions de métaux lourds.

Les résultats ont montré que nous pouvions détecter les ions de métaux lourd avec des limites de détection proches de la ppm : 0,5 ppm pour le cadmium, 2 ppm pour le zinc et le cobalt, 5 ppm pour le nickel et 40 ppm pour le plomb : Les mauvais résultats obtenus pour le plomb étant attribués au fait que cet ion s'adsorberait sur la membrane en BSA. Des tests de stabilité de ce capteur ont permis de montrer qu'après un mois de stockage dans des solutions tampons à 4°C la réponse restait stable. Par ailleurs, on rappelle ici les principaux avantages de ce type de biocapteur par rapport à l'utilisation d'enzymes libres:

(1) une consommation en enzymes immobilisées mille fois moindre

(2) une réduction des interférences par les différents modes d'opération

(3) la pré-incubation n'est pas nécessaire

(4) la durée de la procédure d'analyse est de moins de 5 minutes ;

(5) l'inhibition immédiate de l'AP immobilisée bovine avec les ions des métaux lourds est réversible, puisque l'utilisation de re-activateurs n'est pas nécessaire.

Par conséquent, ce type de biocapteur conductométrique pourrait s'adapter à l'analyse de la pollution totale par les métaux lourds à condition toutefois que leur concentration excède la ppm (application industrielle par exemple).

### 3.1 Introduction

This chapter was submitted as an article in [70].

The application of biosensors for determining toxic compounds is a dynamic trend in sensor research. These sensors seem to be very promising for such a purpose, since analytical systems based on them are simple, rapid and selective. They can be of great use for air and water environmental control, food analysis, medicine and industry, in particular with regard to heavy metal ions, which are known to be harmful pollutants. Their toxicity to the enzymes was explained by their fixation on the thiol groups of proteins [69]. The use of immobilized enzyme probes for enzyme inhibitor determination has been reported [71, 72].

A variety of biosensors for the analysis of heavy metals based on their inhibiting activity towards enzymes has been developed. Among them is the technique proposed by Kukla et al. where the residual activity of covalent coupled urease was measured by means of the capacitance pH-sensitive electrolyte–insulator–semiconductor sensor [73]. Conductometric microelectrodes with cross-linked urease were described by Zhylyak et al. for the determination of $Hg^{2+}$, $Cu^{2+}$, $Cd^{2+}$, $Co^{2+}$, $Pb^{2+}$, $Sr^{2+}$ in water solutions. Planar interdigitated electrodes with urease entrapped in sol-gel was used for the measurements of $Cd^{2+}$ and $Pb^{2+}$ ions with 10 ppm detection limit [74].

A few works used alkaline phosphatase (AP) for biosensor analysis of heavy metals [75]. In the same time AP has good sensitivity to heavy metals, high stability and the large number of commercial available forms [76, 77].

Furthermore conductometric interdigitated thin-film microelectrodes have a small size, high sensitivity and low power consumption [78].

Though, this part of the thesis has been motivated by the curiosity to study pure AP immobilized with bovine serum albumin in glutaraldehyde vapors onto conductometric electrodes. Preceding similar researches by our team concerning, mentioned above, immobilizing & transducing system with AP active algal cells have discovered some weaknesses of such perspective system [3, 6, 7]. Living cells biosensors are very complicated system and sometimes their behavior is rather obscure. Thereby it was decided to study more simple AP enzyme biosensor source with already utilized by us immobilizing & transducing system

Thus the aim of this part was to create a conductometric biosensor for heavy metal ions determination based on immobilized with BSA purified enzyme AP as a sensitive element.

### 3.2 Experimental

### 3.2.1. Materials

AP (EC 3.1.3.1) from bovine intestinal mucosa with activity 24 U/mg, bovine serum albumin (BSA) and 25% aqueous solution of glutaraldehyde (GA) were used for active membrane preparation. $Cd^{2+}$, $Zn^{2+}$, $Co^{2+}$, $Ni^{2+}$, $Pb^{2+}$ nitrates were used as analytes. All chemicals were of analytical grade. All reagents were purchased from Sigma. Substrate solutions of *p*-nitrophenylphosphate disodium salt (pNPP) were prepared immediately before use. All solutions were made with doubly distilled water.

### 3.2.2. Transducer construction

Fig. 3.1 shows the planar conductometric transducers which were fabricated with our recommendations and outlines at the Lashkarev Institute of Semiconductor Physics NASU in Kiev, Ukraine. Two pairs of gold interdigitated electrodes were vacuum-deposited on the pyroceramics substrate (5 x 30 mm). Both the digits width and interdigital distance were 20µm, the sensitive part of each electrode was about 1.5 mm$^2$.

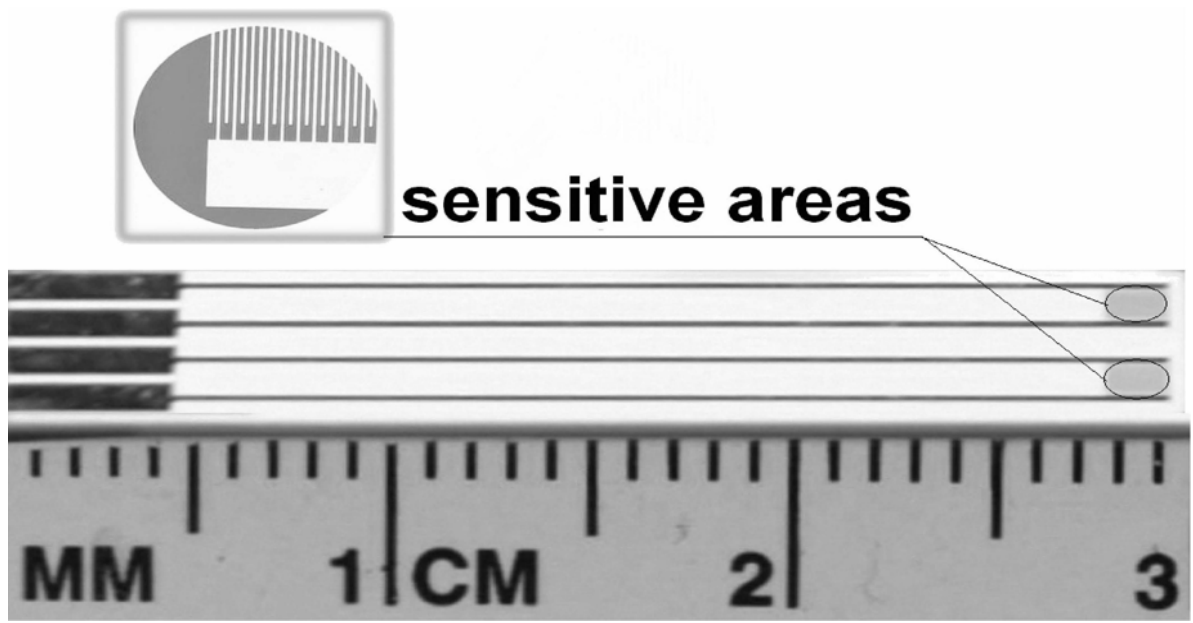

**Fig. 3.1.** *Conductometric micro transducer construction*

### 3.2.3. Enzyme immobilization

The sensitive membrane was formed by crosslinking AP with BSA in saturated GA vapors on the transducer surface during. AP and BSA solutions were prepared in working medium (WM): 10 mM Tris-nitrate, pH 8.6 with 1 mM Mg(NO$_3$)$_2$. Prior to deposition on the sensor chip, these solutions were mixed and glycerol was added (5% AP, 5% BSA and 10% glycerol). The use of glycerol prevents loss of enzyme activity during the immobilization process and results in better membrane homogeneity and adhesion to the transducer surface. The enzyme-containing mixture was deposited on the sensitive surface of the measuring pair of electrodes by the drop method, while the mixture containing 10% BSA and 10% glycerol was deposited on the reference pair of electrodes (Fig. 2.11.). Then the sensor was placed for 30 min in saturated GA vapors. After exposure to GA, the membranes were dried at room temperature for 30 min. Before use the membranes were soaked in WM for at least 30 min to equilibrate the membrane system [79].

### 3.2.4 Measurements

Conductometric measurements were performed by applying to each pair of interdigitated electrodes a small-amplitude alternating voltage 10 mV with 100 kHz frequency generated by a low-frequency wave-form generator G3-112/1, Russia. Output signal after selective preamplifier Unipan-233, Poland, was analyzed using lock-in amplifier Unipan-232B, Poland (Fig. 2.11). The substrate concentration was increased step-wise by adding defined volumes of appropriate concentrated solutions. The conductivity changes result from the enzymatically catalyzed hydrolysis of pNPP. Time to obtain enzymatic signal of described biosensor was less than 1 minute.

The AP inhibition by heavy-metal ions results in a reduction of the biosensor sensitivity to substrate after an immediate (less than 1 min) contact with analyte. Optimization of the biosensor for heavy metal determination was conducted with various substrate concentrations. Measurements were carried out under stirring at room temperature in a 2 ml glass cell filled with WM. Biosensors were prepared each day and stored at 4°C between experiments in WM.

### 3. 3. Results and discussions

The alkaline phosphatase conductometric biosensor is based on the following reaction:

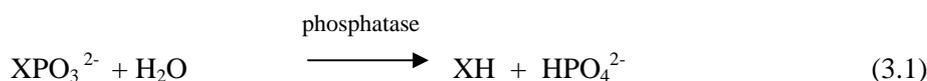

$$XPO_3{}^{2-} + H_2O \xrightarrow{\text{phosphatase}} XH + HPO_4{}^{2-} \qquad (3.1)$$

where $XPO_3{}^{2-}$ is the organic phosphate substrate and $HPO_4{}^{2-}$ the monohydrophosphate.

When paranitrophenyl phosphate is used as a substrate, the product XH is paranitrophenol. Therefore, in the presence of alkaline phosphatase, the reaction induces a change in pH and in conductivity.

The change in conductivity can be detected with a conductometric microtransducer. This electrochemical method based on measuring conductivity change of the analyzed medium. In our case, conductivity change results from enzymatic reaction, enzyme activity, and also depends on the physicochemical properties of reaction medium.

The first step of this work was to study the influence of AP concentrations in membrane on biosensor response to substrate (Fig 3.2).

At low substrate concentration range biosensor with low charged (1 U/ml) membrane has linear and moderate sensitivity to substrate. Whereas high enzyme charged (20 and 70 U/ml) biosensors have bigger amplitude and linear sensitivity range up to 0.2-1 mM pNPP.

At high (1-8 mM pNPP) substrate concentrations high charged sensors have no sensitivity to substrate because of immobilized enzyme saturation by substrate. In the same time low enzyme charged sensor has good signal amplitude and linear range about 1-6 mM pNPP. This phenomena can be explained in following manner, for high concentrations of enzyme, most substrate molecules are consumed at the interface membrane-solution and the product molecules are immediately removed to the bulk solution, so that only a small amount can reach the transducer sensitive area. Consequently, a low signal is observed.

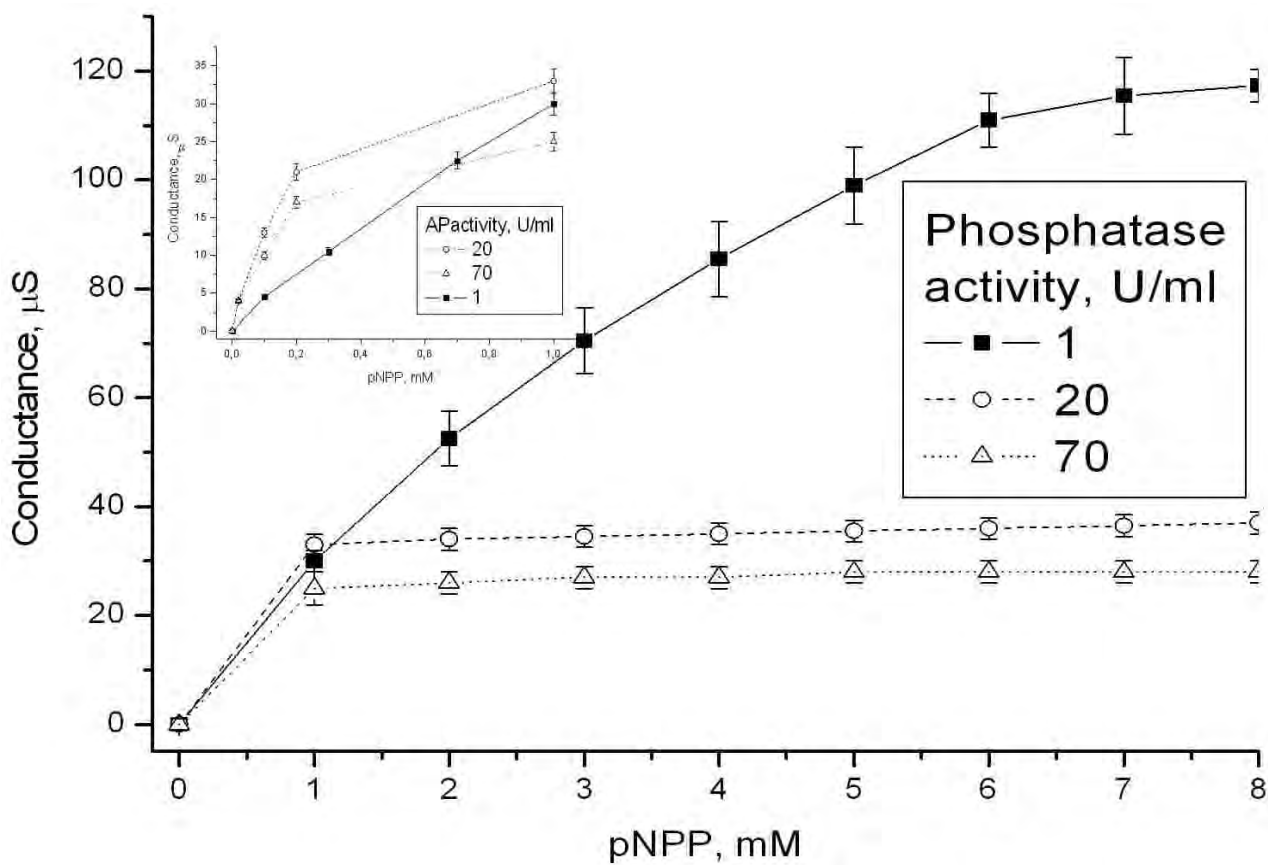

**Fig 3.2.** *Biosensor response to 0.02-1 and 1-8 mM pNPP concentrations as a substrate with different AP concentrations immobilized in BSA gel reticulated with GA. Measurements were performed with 10 microns characteristic dimension platinum conductometric transducer deposited on silicon oxide substrate. Signal frequency and amplitude were 100 kHz and 10 mV respectively. Measurements were conducted in 10 mM Tris-nitrate buffer, pH 8.6 with 1mM Mg(NO₃)₂*

Response amplitudes are much higher for low enzyme concentration. Therefore for high enzyme loaded biosensors (in comparison with low loaded) for small substrate concentrations response amplitudes are bigger, and for high pNPP concentrations they decrease [80].

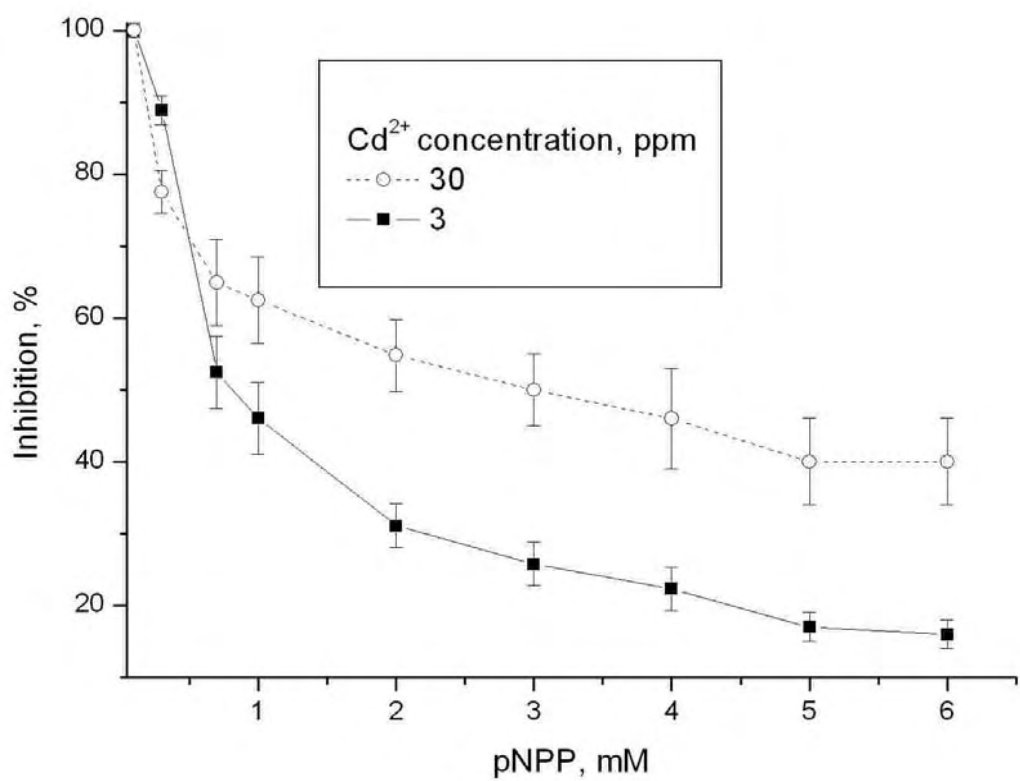

**Fig. 3.3**. *Percentages of AP inhibition for 30 and 3 ppm Cd $^{2+}$ with 2 mM substrate concentration. Measurements were performed with 10 microns characteristic dimension platinum conductometric transducer deposited on silicon oxide substrate. Signal frequency and amplitude were 100 kHz and 10 mV respectively. Measurements were conducted in 10 mM Tris-nitrate buffer, pH 8.6 with 1mM Mg(NO$_3$)$_2$*

Influence of substrate concentrations on biosensor sensitivity to Cd$^{2+}$ concentration was observed (Fig. 3.3.). If the substrate concentration is small, the signal amplitude is too small to be measured properly. If it is higher, the sensitivity to inhibitors decreases. A concentration of 2 mM in pNPP was chosen to prepare our biosensor for heavy metals determination since it provides sufficient stable signal to substrate with a rather good sensitivity to heavy metals.

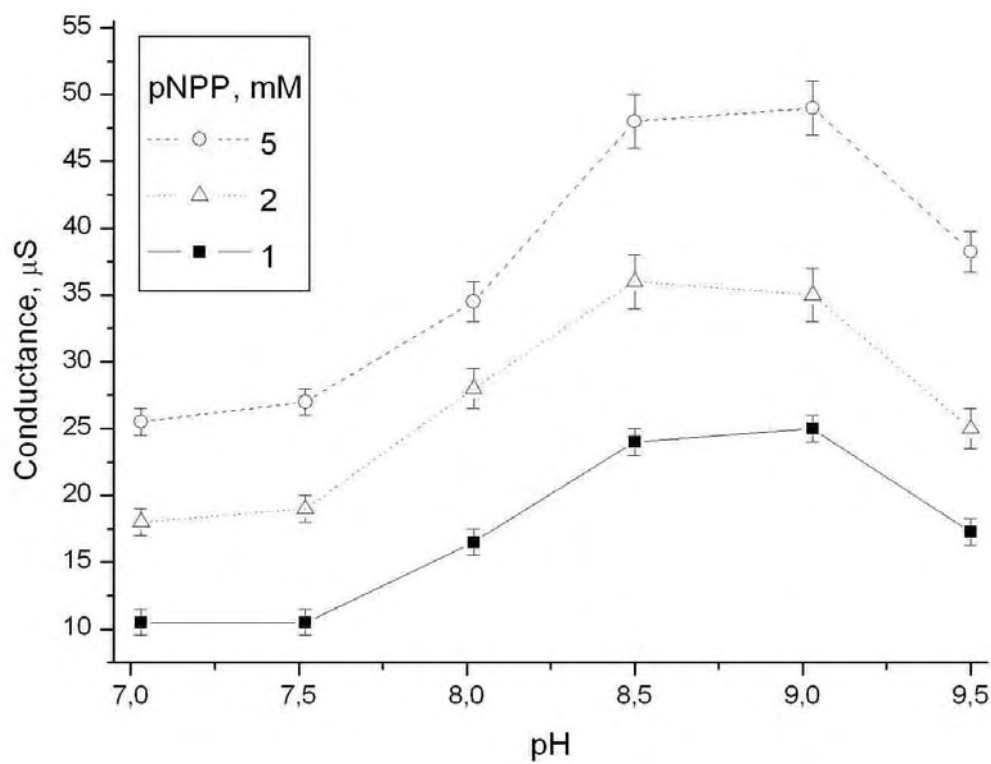

**Fig. 3.4.** *Biosensor response to 5, 2 and 1 mM pNPP concentrations as a function of pH. Measurements were performed with 10 microns characteristic dimension platinum conductometric transducer deposited on silicon oxide substrate. Signal frequency and amplitude were 100 kHz and 10 mV respectively.Measurements were conducted in 10 mM Tris-nitrate buffer, pH 7- 9.5 with 1mM Mg(NO$_3$)$_2$,.*

It is well known that enzyme activity and optimal pH change after immobilization particularly when the enzymatic reaction changes the pH of the medium. Fig. 3.4 shows the dependence of enzyme activity on pH when the enzyme is immobilized on the conductometric transducer. Whatever the substrate concentration, enzyme activity of AP is maximal for pH 8.5-9.0 for our biosensor response to substrate, while it was reported 9.8 for soluble enzyme [76].

Ionic strength is one of the most influential parameters in conductometric assays, since the ionic species, charges and motilities are detected with conductometric measurement.

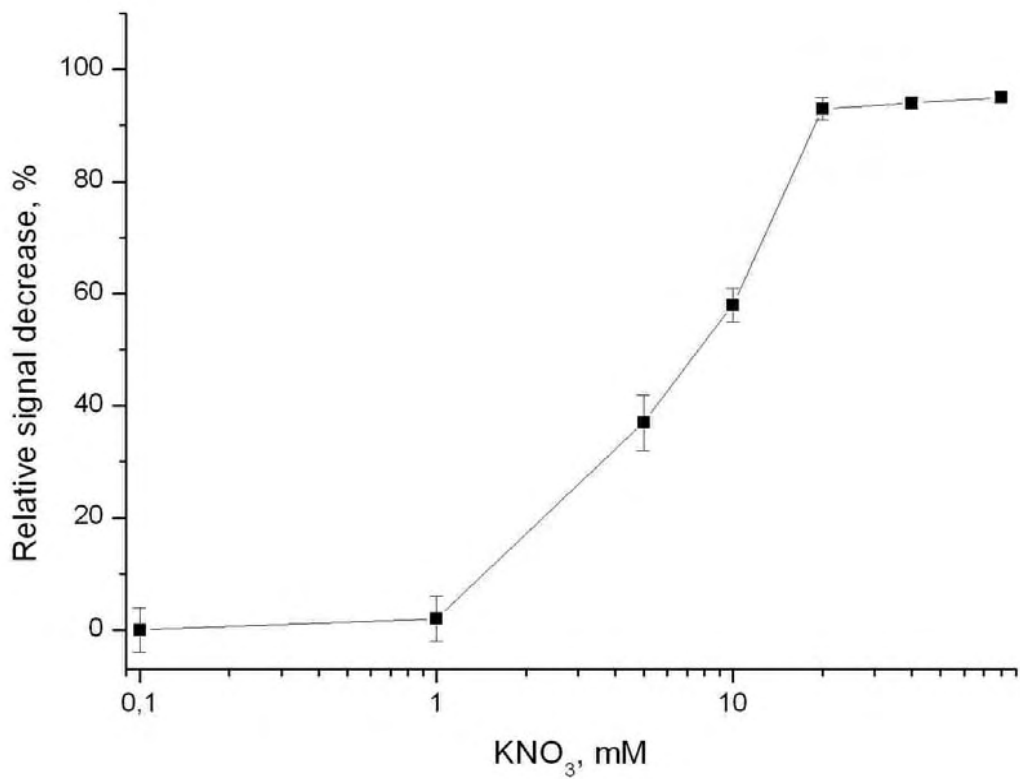

**Fig. 3.5.** *Biosensor relative signal decrease as a function of ionic strength with 2mM substrate concentration . Measurements were performed with 10 microns characteristic dimension platinum conductometric transducer deposited on silicon oxide substrate. Signal frequency and amplitude were 100 kHz and 10 mV respectively. Measurements were conducted in 10 mM Tris-nitrate buffer, pH 8.6 with 1mM Mg(NO_3)_2*

The biosensor relative signal decrease for 2 mM substrate was measured as a function of $KNO_3$ concentrations (Fig. 3.5).

High $KNO_3$ concentrations produce significant background ions interferences and reduce response to substrate by decreasing the amplitude of the enzymatic signal. Sensitivity of the biosensor to ionic strength can be decreased with additional permselective membranes [81].

Fig. 3.6 shows percentages of AP inhibition as a function of various metal ions concentrations. Their relative toxicity toward the enzyme is ranged as follows:

$Cd^{2+} > Co^{2+} > Zn^{2+} > Ni^{2+} > Pb^{2+}$.

At first sight detection limit for $Cd^{2+}$ ions was 0.5 ppm, but only cadmium ions at low concentrations (up to 1 ppm) have some activating influence to immobilized bovine AP. Similar phenomena has been reported in [73] for immobilized urease and $Sn^{2+}$ ions, and also in [82] for immobilized AP and $Ni^{2+}$ or $Co^{2+}$ ions. Probably observed effects are not connected with the enzyme reaction and we observed artifacts. In any case, such phenomena have to be study broader. Thus presently reliable detection limit for $Cd^{2+}$ is about 4 ppm.

For other metals detection limits was about 2 ppm for $Zn^{2+}$ and $Co^{2+}$, 5 ppm for $Ni^{2+}$ and 40 ppm for lead ions. Slight sensor toxicity to $Pb^{2+}$ can be explained by possibility of interactions between lead and BSA molecules [83].

The biosensor practicability is often limited by its operational and storage stability. These characteristics were therefore investigated. Fig. 3.7 shows combined test of operational and storage stability in working conditions for the developed biosensor. In this investigation the reproducibility and operational stability was analyzed during some hours, then the biosensor was stored in the working buffer solution at room temperature for a night, and after this was tested again. This procedure was repeated during 6 days. The responses of the AP conductometric biosensor during several hours were stable and RSD was 4 %. A test of operational stability demonstrated that biosensor responses remain stable with a drift of about 7% per day, when the evolution was monitored during operational conditions.

Furthermore, the storage stability in the WM at 4 ºC was quite good, the biosensor responses remaining stable for more than 1 months.

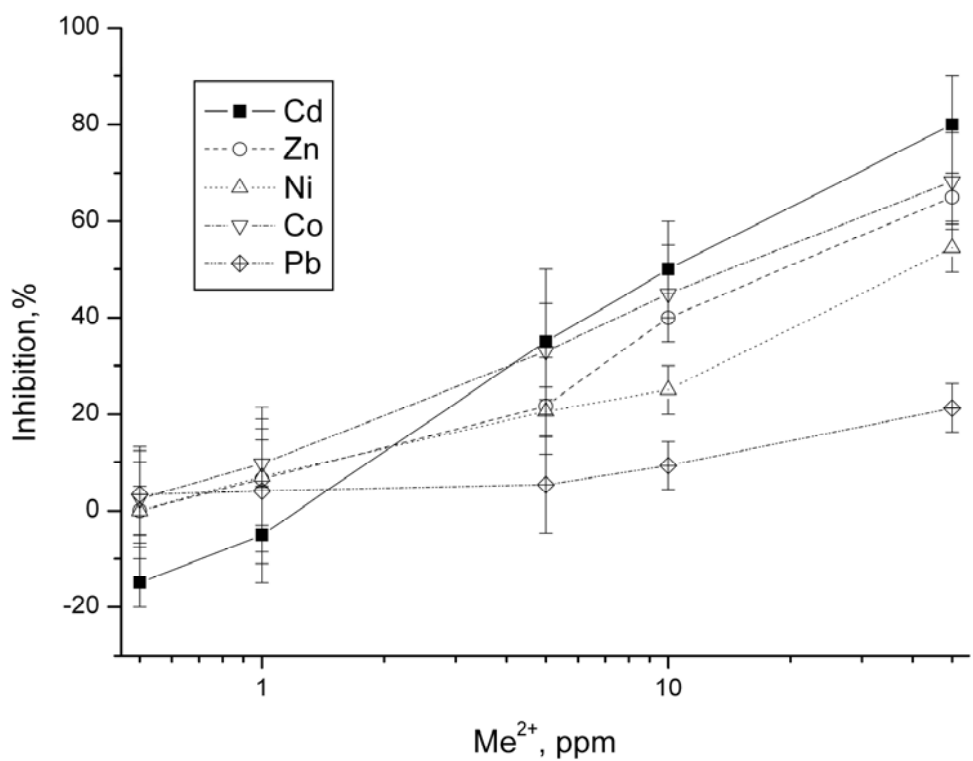

**Fig. 3.6.** *Percentage of AP inhibition as a function of various metal ions concentrations with 2mM substrate concentration. Measurements were performed with 10 microns characteristic dimension platinum conductometric transducer deposited on silicon oxide substrate. Signal frequency and amplitude were 100 kHz and 10 mV respectively. Measurements were conducted in 10 mM Tris-nitrate buffer, pH 8.6 with 1mM Mg(NO₃)₂*

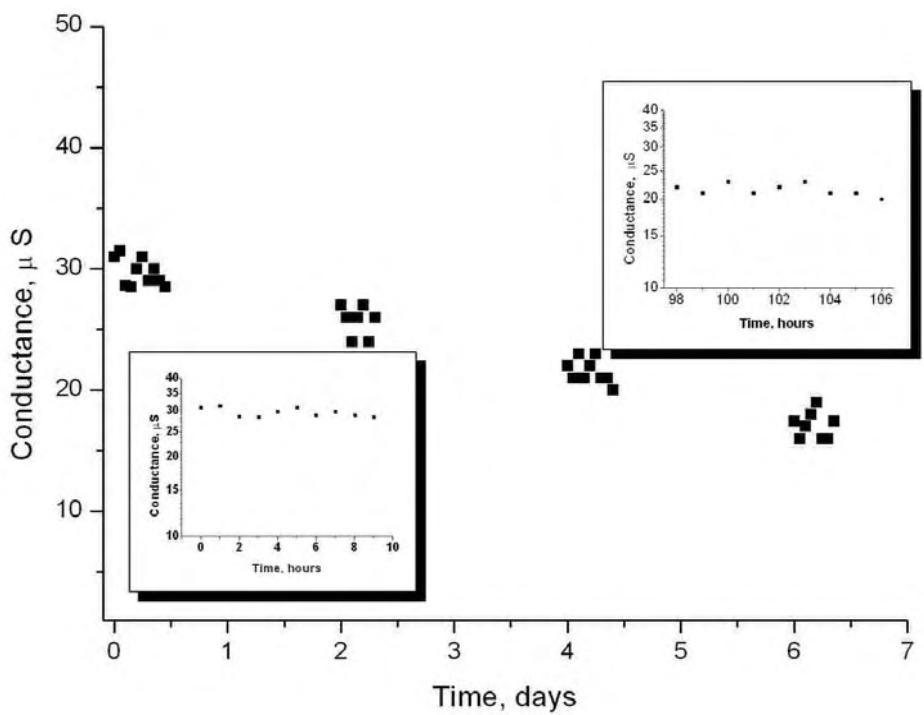

**Fig. 3.7.** *Operational stability of the AP conductometric biosensor after successive assays. Measurements were performed with 10 microns characteristic dimension platinum conductometric transducer deposited on silicon oxide substrate. Signal frequency and amplitude were 100 kHz and 10 mV respectively. Measurements were conducted in 2mM pNPP and 10 mM Tris-nitrate buffer, pH 8.6 with 1mM Mg(NO₃)₂*

### 3.4. Conclusions

AP conductometric biosensors consisting of interdigitated gold electrodes and enzyme membranes formed on their sensitive parts have been used for an estimation of water pollution with heavy-metal ions. The application of immobilized AP has several benefits in comparison with free enzyme [82]: (1) thousands times lower consumption of immobilized enzyme [2, 84]; (2) reduction of interferences by the differential mode of operation ; (3) unnecessary pre-incubation; (4) less than 5 minutes rapid analysis procedure; (5) biosensor requires no re-activators like EDTA. Sensitivity of the biosensor to ionic strength is rather high as for all conductometric sensors and can be decreased with additional membranes [85]. The described conductometric biosensor can be used only for analysis of water pollution by heavy-metal ions due to its low selectivity. Further development will include searching for ways to increase selective detection of heavy-metal ions by using multienzymatic biosensors  and multivariable correspondence analysis [86]

# Chapter 4

**Chapter 4. Conductometric biosensor based on sol-gel entrapped microalgae**

**Résume en français du chapitre 4**

L'utilisation des biocapteurs pour le contrôle de la qualité de l'air et de l'eau constitue un enjeu majeur pour le monitoring des écosystèmes, notamment pour la détection des métaux lourds. C'est dans cet objectif que diverses études ont contribué au développement de capteurs multienzymatiques permettant la détection de différents groupes de polluants. Toutefois ces capteurs sont généralement peu spécifiques et sensibles pour chacune des activités mesurées. L'utilisation d'enzymes isolées comme biorécepteurs leur confère, d'autre part, une stabilité très réduite dans le temps.

C'est pour palier à ces différents inconvénients que des biocapteurs enzymatiques utilisant des organismes unicellulaires comme biorécepteurs ont été développés. Ils sont basés sur la mesure de l'activité d'une ou plusieurs enzymes membranaires du biorécepteur. Leur réponse possède ainsi une signification écologique importante puisque les enzymes opèrent dans leur environnement cellulaire naturel. Dans des études antérieures, nous avons développé des capteurs à cellules algales utilisant la microalgue *Chlorella vulgaris* immobilisée par différentes techniques (gel d'albumine de sérum de bœuf réticulé au glutaraldehyde, gels d'agarose ou d'alginate de calcium). Ces différents gels constituent des barrières entre le biorécepteur et le milieu réactionnel pouvant limiter l'accès des cellules algales aux molécules recherchées. Ces procédés d'immobilisation risquent donc de réduire la sensibilité du biocapteur de manière significative. L'étude présentée ici, propose un biocapteur à cellules algales immobilisées dans une matrice de silice (technique sol-gel) sur une microélectrode conductimétrique. Ce biocapteur est basé sur la mesure de l'activité phosphatase alcaline membranaire des algues immobilisées, dans l'objectif de détecter la présence de métaux lourds, ces derniers s'étant révélés inhibiteurs de l'activité phosphatase dans des études antérieures. Le principal intérêt des matrices de silice est d'être très peu réactives vis-à-vis de la majorité des composés chimiques. Elles sont d'autre part très peu coûteuses.

Le mode opératoire consiste à ajouter une suspension algale à une solution de silicate de sodium juste avant sa polymérisation induite par une acidification du milieu. Le mélange est alors déposé sur la partie sensible d'électrodes interdigitées en or au contact de laquelle il se solidifie. La cinétique de l'activité de la phosphatase alcaline est réalisée en ajoutant des quantités croissantes de para-nitrophényl phosphate (substrat de la réaction) au contact de l'électrode. Les résultats montrent qu'une activité optimum est obtenue pour une concentration en cellules algales de $10^8$ cell/ml immobilisées dans la membrane sol-gel. La réponse du biocapteur est optimale pour un pH de 8,5. Des Limites de détection de l'ordre de 1 ppb ont été obtenues pour $Cd^{2+}$, $Co^{2+}$, $Ni^{2+}$, $Pb^{2+}$ et de 10 ppb pour $Zn^{2+}$. Des mesures d'activité au cours du temps mettent en évidence une stabilité du capteur d'au moins 40 jours  Enfin, les résultats montrent qu'une augmentation de la force ionique du milieu réactionnel induit rapidement des perturbations de la réponse du capteur. Ce développement devra être poursuivi dans l'objectif d'améliorer la sensibilité du capteur ainsi que sa sélectivité vis-à-vis des métaux lourds.

### 4.1. Introduction

This chapter was accepted as an article in [87].

The application of biosensors for determination of toxic compounds is a dynamic trend in sensor research. These sensors seem to be very promising since analytical systems based on them are simple, rapid, and selective. They can be of great use for air and water environmental control, food analysis, medicine and industry, in particular as regards to heavy metal ions, which are known to be harmful pollutants. Heavy-metal ions toxicity to living organisms was explained by their fixation on the thiol groups of enzymes. This is the case of alkaline phosphatase which is known to be inhibited by this kind of pollutants [69].

The development of a multi-enzymatic biosensor for the detection of different groups of pollutants represents an important challenge. However, a biosensor using different enzymes on a multisensor array cannot operate in optimal conditions, since they may be different from one enzyme to the other. Other problems including enzyme stability and enzyme purification cost must also be overcome. The use of whole cells or microorganisms to produce a multi-enzymatic biosensor can be a good solution, since they contain a large number of enzymes. Moreover, ecotoxicological information can be obtained from the effects of pollutants on these living organisms.

Among the tested immobilizing techniques used by our team there are glutaraldehyde cross linking, calcium alginate and agarose gel entrapments, pyrrol-alginate gel electropolymerization [6, 7, 88]. All these methods are unsuitable for our application because of physical and chemical instability and/or back side reactions.

Silica matrixes are relatively inexpensive to synthesize and have interesting properties including biocompatibility and chemical inertness [89]. Special features of such systems are great possibilities of the variation of the physical, chemical, and functional properties of materials with the identical or close composition of the reaction products. Sol-gel based membranes also reduce side reactions compared to conventional supports.

In this communication, the sol-gel technique has been used to construct a conductometric biosensor based on thin-film planar interdigitated microelectrodes. The main benefits of such microsensors are small size, high sensitivity, and low power consumption [78].

Thus, the aim of our work was to create a conductometric biosensor for heavy metal ions determination based on entrapped in sol-gel whole cells of *Chlorella vulgaris* as a sensitive element.

### 4.2. Experimental

### 4.2.1. Chemicals

Silica sources were sodium silicate solutions (purchased from *Riedel-de-Haen*) and colloidal silica LUDOX HS-40 (from *Aldrich*). All other reagents were purchased from *Sigma*. Zn, Cd, Co, Ni, and Pb nitrates were used as analytes. All chemicals were of analytical grade. Alkaline phosphatase substrate solutions of p-nitrophenylphosphate disodium salt (pNPP) were prepared immediately before use.

### 4.2.2. Cell culture

The *Chlorella vulgaris* strain (CCAP 211/12) was purchased from The Culture Collection of Algae and Protozoa at Cumbria, United Kingdom. The axenic algal strain was grown in the culture medium and under conditions described by the International Organization for Standardization (ISO 8692) [90].

### 4.2.3. Sensor design

Fig. 3.1 shows the planar conductometric transducer. Two identical pairs of gold interdigitated electrodes (thickness 0.5 µm dimensions 5 x 30 mm) were fabricated by vacuum deposition on a ceramic substrate (sintered aluminum oxide) at the Institute of Semiconductor Physics, Kyiv, Ukraine. An intermediate layer of chromium (0.1 µm thick) was used for better gold adhesion. Each finger of the electrode was 20 µm wide and 1 mm long, with 20 µm spacing between fingers of the electrode in the pair. The sensitive area of each electrodes pair was about 1 x 1.5 mm. To define the sensitive area of the transducer, the central part of the chip was covered with epoxy resin.

### 4.2.4. Cell entrapment

Sodium silicate (0.4 M, 4 ml) and colloidal silica LUDOX (8.5 M, 4 ml) were thoroughly mixed (300 rpm) to obtain a homogeneous silica solution. An HCl, 4 M solution was then added drop by drop until an appropriate pH was reached to induce the gelation process. Before gelation, an algal suspension containing $3 \times 10^8$ cells/ml and 10% (w/w) glycerol was introduced under stirring. The resulting solution was deposited on the sensitive surface of the measuring pair of electrodes by the drop method (0.15 µl) to produce silica matrixes containing algal cells. The measuring pair of electrodes was covered with an AP active algal membrane, while the reference pair used algal cells with no AP activity. Gelation occurred within about 5 min at room temperature. Wet gels were aged for 24 hours at 4°C in the mother solution in a closed flask in order to ensure gel densification before analysis [91].

### 4.2.5 Measurements

Conductometric measurements were performed by applying to each pair of interdigitated electrodes a small-amplitude alternating voltage 10 mV with 100 kHz frequency generated and analyzed by SR-830 DSP lock-in amplifier, Stanford research systems, USA (Fig. 2.11) [92]. The substrate concentration was increased step-wise by adding defined volumes of appropriate concentrated solutions. The conductivity changes resulted from the enzymatically catalyzed hydrolysis of pNPP. The AP inhibition by heavy-metal ions resulted in a reduction of the biosensor sensitivity to substrate. All measurements were carried out under stirring at room temperature in a 2 ml glass cell filled with working medium: 1 mM $Mg(NO_3)_2$, as AP activator, with 10 mM Tris-nitrate buffer, pH 8.5 [93]. Biosensors were prepared each day and stored at 4°C between experiments. Storage was conducted in the culture medium without phosphate ions to avoid AP activity loss and cells growth [94, 95].

**4.3 Results and discussions**

Described algal conductometric biosensor is based on the following reaction:

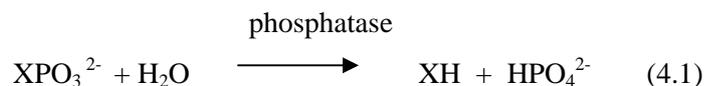

$$XPO_3{}^{2-} + H_2O \xrightarrow{\text{phosphatase}} XH + HPO_4{}^{2-} \quad (4.1)$$

where XPO3 2- is the organic phosphate substrate and HPO42- – the monohydrophosphate.

When paranitrophenyl phosphate is used as a substrate, the product XH is paranitrophenol.

Therefore, in the presence of alkaline phosphatase, the reaction induces a change in pH and in conductivity. The change in conductivity can be detected with a conductometric microtransducer. This electrochemical method is based on measuring conductivity change of the analyzed medium. In our case, conductivity change results from enzymatic reaction, enzyme activity, and also depends on the physical and chemical properties of reaction medium.

The first step of this work was to optimize the biosensor response as a function of algal concentrations in the membrane (Fig. 4.1).

As for the algae concentrations in the membranes of described biosensor, there is the optimal concentration *i.e.* $(22{\div}640){\times}10^{6}$ cells/ml. It was observed that the signal amplitude decreased at higher algae concentrations. In this case enzymatic reactions can only occur on the border of the membranes preventing substrate molecules from diffusing inside and reacting with algae situated near the sensitive areas. As a consequence, a low signal was observed. Moreover, lower algae concentrations in membranes also give slight conductivity variations since only a few substrate molecules can be transformed. It is interesting to note that for enzyme biosensors the same conclusion has already been done [80].

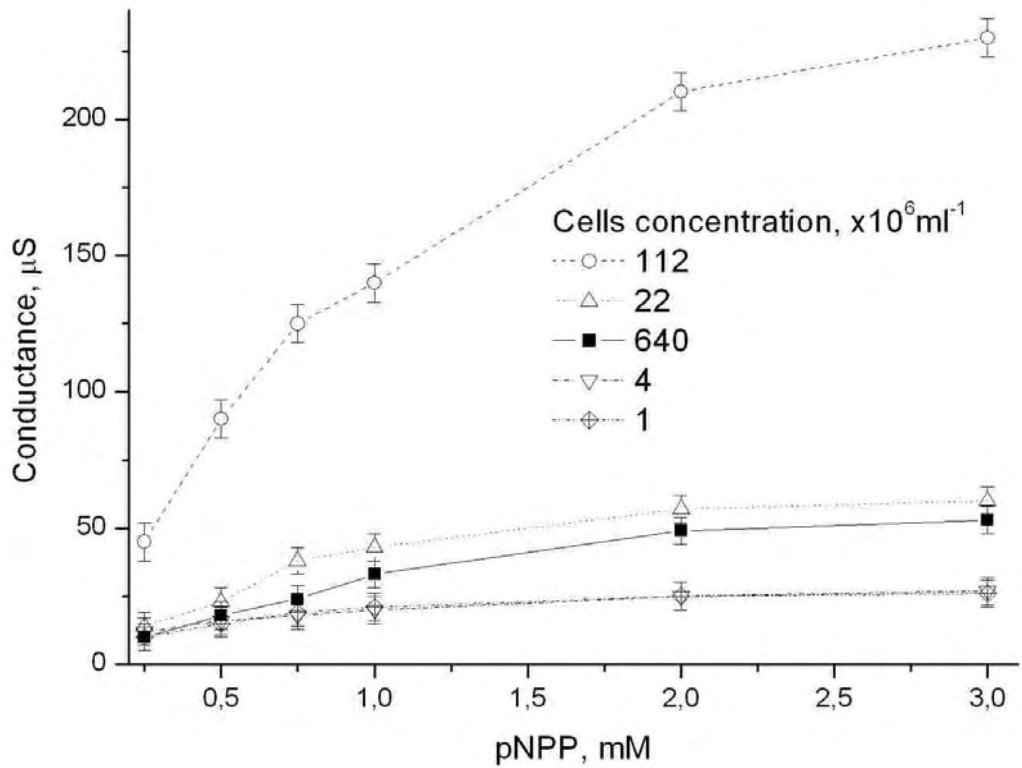

**Fig. 4.1.** *Dependence of conductometric biosensor response on substrate concentration in working medium for different algae concentrations in sol gel membranes. Measurements were performed with 10 microns characteristic dimension platinum conductometric transducer deposited on silicon oxide substrate. Signal frequency and amplitude were 100 kHz and 10 mV respectively. Measurements were conducted in 10 mM Tris-nitrate buffer, pH 8.6 with 1mM $Mg(NO_3)_2$.*

It's important to note, concentration of algal cells is not only one influence of final enzymatic activity of biosensor. Specific algal suspension activity should be controlled because its variability between the cultivations is possible. It is well known that enzyme activity changes after immobilization because a bundle of different reasons, for example, through protein three-dimensional structure changes or because of manual membrane deposition mode. Such activity changes are different from one biosensor to another even in the same biosensor preparation series. It's almost impossible in such preliminary stage of biosensor creation to obtain sensor with precisely defined enzymatic activity. Above all, we don't need this echelon of accuracy. After deposition all the sensors should be tested and calibrated and we can choose the most optimal samples for our following measurements. By the similar reasons it seems as unnecessary to compare absolute enzymatic membrane activity between described enzymatic and algal AP biosensors.

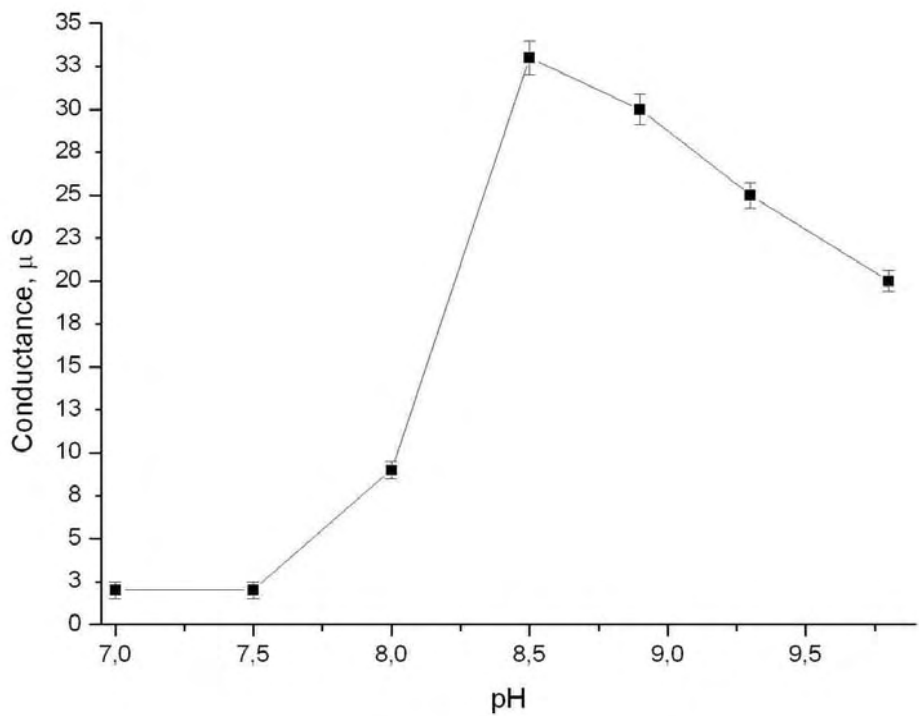

**Fig. 4.2.** *The pH-Dependence of conductometric biosensor response for 2 mM of pNPP. Measurements were performed with 10 microns characteristic dimension platinum conductometric transducer deposited on silicon oxide substrate. Signal frequency and amplitude were 100 kHz and 10 mV respectively. Measurements were conducted in 10 mM Tris-nitrate buffer, pH 8.6 with 1mM Mg(NO₃)₂*

Influence of substrate concentrations on biosensor sensitivity to $Cd^{2+}$ concentration has already been observed. If the substrate concentration is small, the signal amplitude is too small to be measured properly. If it is higher, the sensitivity to inhibitors decreases. The concentration of 2 mM in pNPP was chosen to prepare our biosensor for heavy metals determination since it provides the sufficient stable signal to substrate with rather a good sensitivity to heavy metals [96].

As it is well known that enzyme activity and optimal pH are changed after immobilization particularly when the enzymatic reaction changes the pH of the medium.

Fig. 4.2 shows the dependence of algal AP activity on pH when the cells are immobilized on the conductometric transducer. Enzyme activity of *Chlorella vulgaris* is maximal for pH 8.5 for our biosensor response to substrate, while it was reported 10.5 for the algae in suspension [93].

Ionic strength is one of the most influential parameters in conductometric assays, since the ionic species, charges and motilities are detected using conductometric measurements. The biosensor response to 2 mM substrate was measured as a function of $KNO_3$ concentrations (Fig. 4.3).

High $KNO_3$ concentrations produce significant background ions interferences and reduce response to pNPP by decreasing the amplitude of the enzymatic signal.

The biosensor is relatively stable for 40 days under storage conditions (Fig.4.4).

Good correlation in terms of AP storage stability was found between the biosensor and the cells suspension [93].

Fig. 4.5 shows the percentage of AP inhibition as a function of various metal ions concentrations. Inhibition of AP was found for ppm concentration levels of tested metals while activation of AP activity occurs at ppb concentration levels, except for Zn. The activation could be explained by cellular stress: indeed, to prevent the cell from heavy-metals damages, stress promoters are produced inducing an increase of some enzymatic activities [97]. By the same reason algal biosensor is more sensitive than purified AP based device. Moreover living cell provides optimal conditions for its enzymes. Such natural conditions holds by homeostasis are impossible artificially create for purified enzyme biosensor.

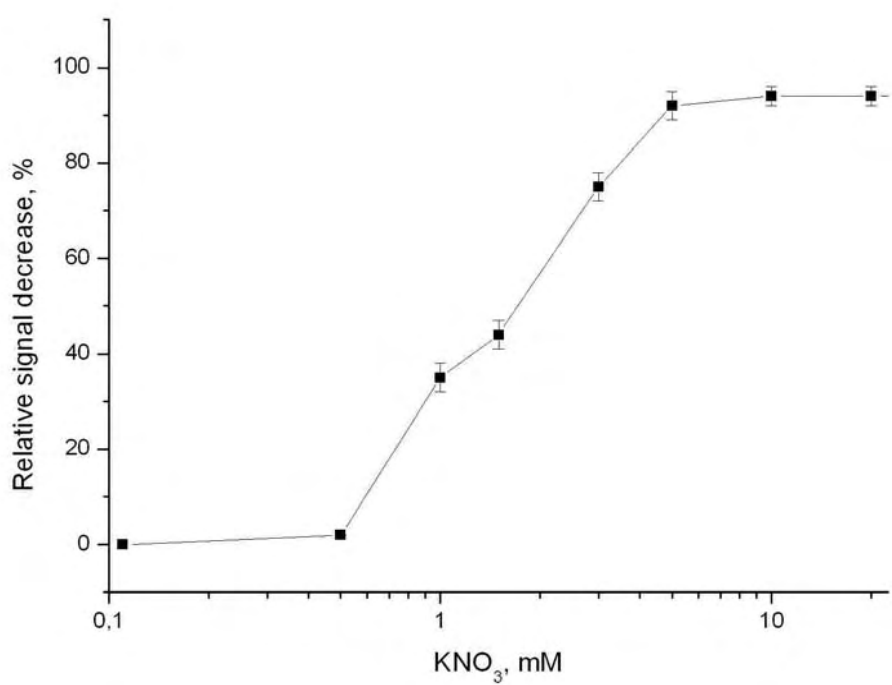

**Fig. 4.3.** *Dependence of conductometric biosensor relative signal decrease on ionic strength for 2 mM pNPP. Measurements were performed with 10 microns characteristic dimension platinum conductometric transducer deposited on silicon oxide substrate. Signal frequency and amplitude were 100 kHz and 10 mV respectively. Measurements were conducted in 10 mM Tris-nitrate buffer, pH 8.6 with 1mM Mg(NO₃)₂.*

Making the measurements with described biosensor it can be finding that inhibition percentage of - 20% corresponds to 2 inhibitor concentrations 1 ppb and 20 ppb. Practically to refine such type of data it is advisable to work with sample dilution series and to analyze inhibition levels obtained from these dilution series.

It has been shown there are no effects of antagonism and synergism for algal sensor at working concentrations of cadmium and zinc ions [3]. More detailed investigations of different heavy metals mixtures by described biosensor are among the research perspectives.

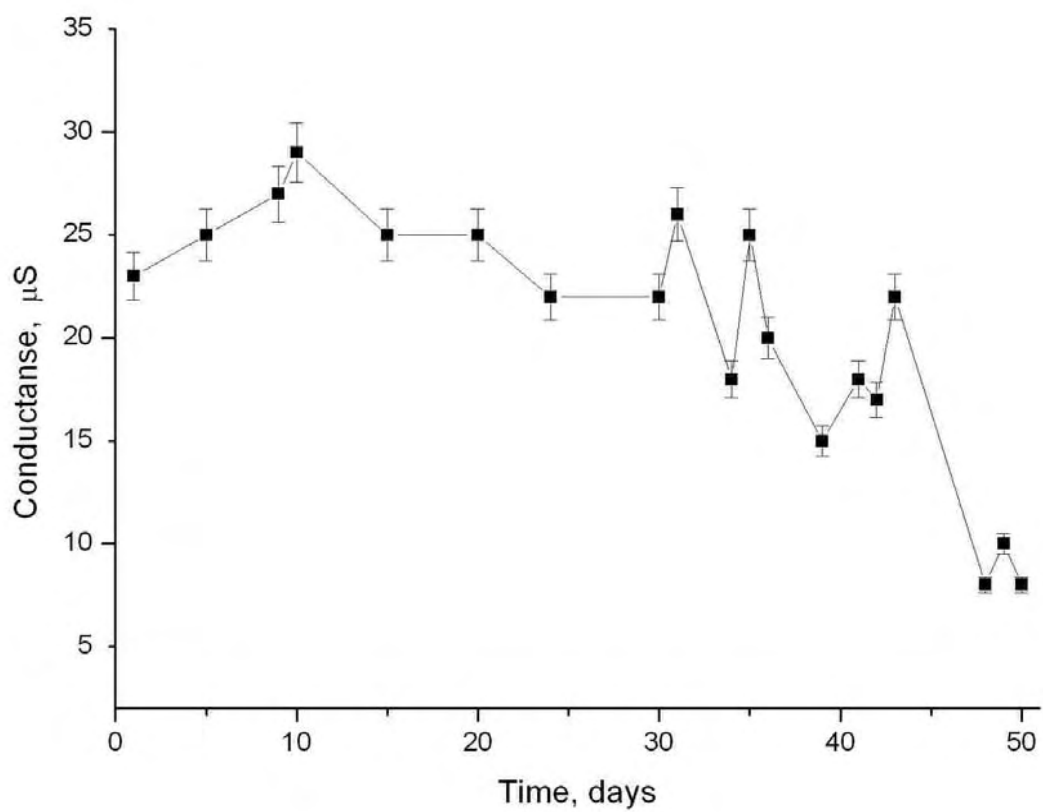

**Fig. 4.4.** *Dependence of the algal biosensor response on storage life for 2 mM pNPP. Measurements were performed with 10 microns characteristic dimension platinum conductometric transducer deposited on silicon oxide substrate. Signal frequency and amplitude were 100 kHz and 10 mV respectively. Biosensor was stored between measurements in phosphateless nutrient medium at 4ºC. Measurements were conducted in 10 mM Tris-nitrate buffer, pH 8.6 with 1mM Mg(NO₃)₂.*

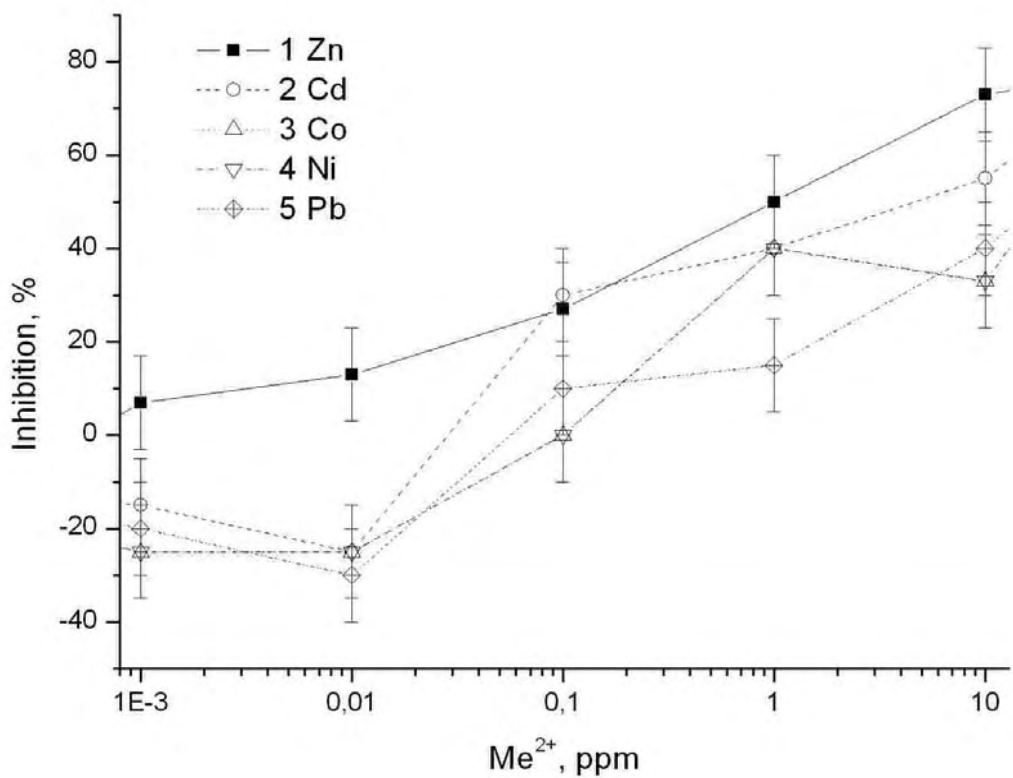

**Fig. 4.5.** *Calibration curves of conductometric algal biosensor for different metal ions. Measurements were performed with 10 microns characteristic dimension platinum conductometric transducer deposited on silicon oxide substrate. Signal frequency and amplitude were 100 kHz and 10 mV respectively. Measurements were conducted in 10 mM Tris-nitrate buffer, pH 8.6 with 1mM Mg(NO₃)₂*

### 4.4. Conclusions

AP conductometric biosensors consisting of interdigitated gold electrodes and algae entrapped in sol-gel membranes formed on their sensitive parts have been used for the assessment of water pollution with heavy metals ions.

Optimal algal concentrations in the membranes of the described biosensor were about $(30 \div 300) \times 10^6$ cells/ml. Optimal pH of the biosensors working medium was 8.5.

Detection limits were about 1 ppb for $Cd^{2+}$, $Co^{2+}$, $Ni^{2+}$, $Pb^{2+}$ and 10 ppb for $Zn^{2+}$. The storage stability of the biosensor was more than 40 days.

The sensitivity of the biosensor to ionic strength is rather high for all conductometric sensors and can be decreased with additional permselective membranes [98]. Further development will include searching for ways to improve selectivity to heavy metal ions using multi-enzymatic biosensors arrays and multivariable correspondence analysis [86].

# Chapter   5

**Chapitre 5: Application of the conductometric biosensor for an assessment of heavy metal ions in urban waters**

**Résume en français du chapitre 5**

La gestion durable des eaux urbaines nécessite des outils d'alarme précoce permettant de détecter la présence de composés toxiques de manière à réagir rapidement en cas de pollution. Les biocapteurs peuvent remplir ce rôle dans la détection des polluants grâce à leur sensibilité, leur faible coût et leur adaptation aisée à un mode de surveillance *in situ* et en continu. Le fonctionnement des biocapteurs à cellules algales présentés ici est basé sur la perturbation métabolique de cellules immobilisées après exposition à des polluants. Dans ce chapitre, des mesures ont été réalisées au moyen d'un biocapteur conductimétrique sur des eaux pluviales urbaines dans l'objectif d'évaluer la potentialité du biocapteur pour la surveillance de ce type d'effluents. Les résultats sont comparés à ceux obtenus en microplaques 96 puits sur algues libres préalablement exposées aux eaux pluviales.

Le biocapteur est réalisé par immobilisation de cellules de *Chlorella vulgaris* par la technique sol-gel selon le protocole décrit dans le chapitre précédent. Les bioessais sur algues libres sont réalisés sur la même souche algale que celle utilisée dans la conception du biocapteur. Dans les deux cas l'activité des phosphatases alcalines membranaires est mesurée, le substrat utilisé est le 4-para-nitro-phényl-phosphate pour le capteur conductimétrique et le méthyl-umbelliferryl phosphate pour les bioessais sur algues libres. Les résultats expriment le pourcentage d'activité résiduelle après exposition aux eaux pluviales rapporté à l'activité témoin avant exposition. Les résultats montrent que le biocapteur permet de détecter une activité dépendante de la concentration en substrat selon une cinétique de type Michaelis-Menten superposable à celle obtenue avec les algues libres. Le biocapteur a été préalablement calibré avec des solutions d'ions métalliques (Cd, Zn, Pb, Co, Ni). Ces derniers produisent une inhibition de l'activité phosphatase pouvant atteindre 80% du témoin pour une concentration de 100 mg/l. A l'opposé, les plus faibles concentrations (0.1-10 µg/l) produisent une activation de l'activité phosphatase. Après cette étape de calibration des mesures sont réalisées sur des eaux pluviales récoltées en sortie de déversoir d'orage lors d'un épisode pluvieux. Les mesures sont réalisées pour différents pourcentages de dilution des eaux pluviales. Les plus faibles concentrations en eaux pluviales produisent une stimulation de l'activité phosphatase alors que les plus fortes concentrations produisent une inhibition pouvant atteindre 80% de l'activité du témoin. L'analyse physico-chimique des échantillons réalisée préalablement à nos mesures met en évidence des concentrations en cations métalliques pouvant expliquer les résultats obtenus. Les résultats obtenus sur algues libres suivent les mêmes tendances que ceux obtenus en biocapteurs. Toutefois les inhibitions observées sont moins fortes. Il semblerait donc que les biocapteurs soient plus sensibles pour la détection de composés inhibiteurs. Ces résultats semblent à première vue tout à fait cohérent en tenant compte du fait que la concentration algale immobilisée sur le biocapteur est beaucoup plus faible que celle fixée par le protocole des bioessais en microplaques. Le ratio algues / polluants va donc en faveur d'une plus grande quantité de polluants par cellules algale dans le cas des biocapteurs. Des essais standardisés d'inhibition de croissance ont été réalisés simultanément à nos mesures. Ils ne montrent qu'un très faible impact des eaux pluviales. Ainsi le biocapteur conductimétrique testé dans cette étude semble présenter un réel intérêt pour la détection de toxique dans les eaux pluviales. Il devrait pouvoir à terme être implanté sur les sites à surveiller comme outil d'alarme précoce de contamination du milieu.

## 5.1 Introduction

This chapter was presented at [99].

The environmental impact of urban waters (waste and storm) poses a great challenge to environmental protection and the pursuit of sustainable development of urban areas. Their harmful effects on ecosystems were shown in numerous studies [100].The necessity of a sustainable management of urban waters leads to the requirement of EWS to detect different chemicals in situ at very low concentrations in order to react quickly for limiting impact on natural surface and ground waters. Biosensors for pollutants determination can act as EWS thanks for their unique characteristics which include their sensitivity , their low cost and their easy adaptation for continuous detection on-site monitoring [101].

A biosensor can be considered as a combination of a bioreceptor, the biological component, and a transducer, the detection method. The total effect of a biosensor is to transform a biological event into an electric signal. The first link of a biosensor is the bioreceptor, which has a particularly selective site that identifies the analyte and ensures molecular recognition. In this work, bioreceptor is *Chlorella vulgaris* chosen among unicellular algal species because they are present ubiquitously in natural waters and are able to metabolize a wide range of chemical compounds [102].

These whole-cell algal biosensors are based on metabolic perturbations of immobilized cells in the presence of toxicants and have the ability to detect different group of pollutants provided they affect a particular alga metabolic pathway. This is the case of pesticides and heavy metals which are strong inhibitors of acethylcholinesterase and alkaline phosphatase, both are located on *Chlorella vulgaris* membranes.

Compared to biosensors using purified enzyme, whole cell biosensor are more resistant to the activity loss because their enzymes and cofactors are hosted in an environment optimized by nature. Therefore, these biosensors are more suitable to meet all the requirements for environmental surveillance [103]. They can identify in situ the presence of a toxic compound as soon as it released in waste water or aquatic environment. Other whole cell biosensors were constructed from genetically modified cells [104]. Those techniques may improve the biosensor sensitivity and selectivity but are no longer able to reflect the ecosystem operating conditions. In the present work, only native cells have been used to preserve the ecological aspect of the media under study.

Previous studies showed a good correlation between measurements carried out on river waters, polluted by heavy metals and pesticides, and results from chemical analysis [7].

The aim of this chapter is to assess the potentiality of a conductometric biosensor to monitor urban water toxicity. Experiments were conducted on urban water brought to the laboratory. Results were compared with measurements by bioassays.

## 5.2 Methods

### 5.2.1 Cells culture

The *Chlorella vulgaris* strain (CCAP 211 / 12) was purchased from the culture collection of Algae and Protozoa at Cumbria, United Kingdom. The axenic algal strain was grown in the culture medium and under conditions described by the International Organization for Standardization (ISO 8692) [90].

### 5.2.2 Biosensor design

The conductometric transducers were fabricated at the Institute of Chemo- and Bio-sensors (Munster, Germany) [98]. Two pairs of Pt (150 nm thick) interdigitated electrodes were made by the lift-off process on the pyrex glass substrate. The Ti intermediate layer of 50 nm thick was used to improve adhesion of Pt to substrate. Central part of the sensor chip was passivated by $Si_3N_4$ layer to define the electrodes working area. Both the digits width and interdigital distance were 10 µm and their length was about 1 mm. Thus, the sensitive part of each electrode was about 1 $mm^2$ ( Fig. 2.4.).

Measurements are based on the detection of solution conductivity variation inside algal cells immobilized. Alkaline phosphatase induces catalytic reaction consuming / producing different ionic species resulting in measurable conductivity changes.

### 5.2.3. Algae immobilization

Sodium silicate (0,4 M, 4 ml) and LUDOX (8,5 M, 4 ml) were thoroughly mixed to obtain a homogeneous silica solution. An HCl 4 M solution was then added drop by drop until an appropriate pH is reached to induce the gelation process. Immediately an algal solution containing 1.3 x $10^8$ cells / ml and 10 % (w/w) glycerol was introduced and the mixture was deposited on the sensitive area of the electrode.

### 5.2.4. Measurements

#### 5.2.4.1. Enzymatic reaction measurements

For biosensors, measurements were carried out in daylight at room temperature in a 2 ml glass cell filled with Tris-HCl (10 mM, pH 8.4)

Biosensors were immersed in this vigorously stirred solution. After stabilization of the output signal, different aliquots of the substrate stock solution were added into the vessel. The differential output signal (dS) was registered using a "home made" conductometric laboratory apparatus and the steady state response of the biosensor was plotted against the substrate concentration

For bioassays, free algae are used and measurements in microplates are based on fluorescence detected by a spectrofluorometer (Fluostar BMG). The alkaline phosphatase enzymatic reaction using methyl umbelliferyl phosphate (MUP, Sigma) as substrate gives fluorescent product methyl umbelliferone (MUF). Essays were carried out in 96 wells microplates.

Wastewater, collected in Chevire (France), was sterilized before experiment at 130°C, 1.5 bar to suppress contaminating bacterial phosphatase activity.

### 5.2.4.2. Toxicity measurements

For biosensors, dS was measured for a definite substrate concentration. The biosensor was then preincubated in a test solution for 15 mn . After washing dS before and after exposure to the test solution was compared and the residual activity rate was calculated.

For bioassays, 48 wells microplate were filled with algal solution. After sedimentation, culture medium could be removed and replaced by the test solution. Exposure lasts two hours. After removing the test solution and resuspending algae in distilled water, enzymatic activity was measured in 96 wells microplate as above-mentioned.

### 5.3. Results and discussion

### 5.3.1. Enzymatic activity detection using conductometric biosensors

In a previous work, it has been proved that phosphatase alkaline activity can be monitored for immobilized Chlorella vulgaris using conductometric biosensors [7]. The enzymatic activity follows a classical Michaelis-Menten behaviour. The relative standard deviation of the sensor did not exceed 8 %.

### 5.3.2. Biosensor calibration for heavy metal detection

Alkaline phosphatase measurements were carried out before and after exposure to heavy metals ions. Fig 4.5 shows phosphatase activity (expressed as percent of control before exposure). Inhibition can reach 80 % of control for 100 mg/ L heavy metal. At low concentrations (0.1 - 10 µg/l) metal ions as Cd, Pb, Co, Ni enhance phosphatase activity. The activation of some enzymes by low metal concentrations has already been studied and could be explained by cellular stress: indeed to prevent the cell from metal damages, stress promoters are produced inducing an increase of some enzymatic activities [105].

### 5.3.3. Biosensor responses after exposure to urban stormwaters

The biosensor developed is supposed to be used for the detection of pollutants especially heavy metal ions.

Measurements were carried out after exposures to different concentrations of urban stormwater in the mixture assay. Stormwater was sterilized before experiment at 130°C, 1.5 bar for suppressing bacterial phosphatase activity. In Fig. 5.1, biosensor response can be compared with results of Fig 4.5. The lowest concentrations of stormwater enhance phosphates activity as higher concentrations inhibit this activity.

Heavy metal contain of stormwater tested was determinated by chemical analysis (results showed in Table 5.1). A mixture of metal ions with a concentration between ppb and ppm, can explain results obtained.

For the lowest concentrations of stormwater tested an activation of phosphatase activity was obtained due to the low concentration of metal ions. For higher concentrations a good correlation was obtained between concentration of sample (obtained from reference method) and inhibition rate. It should be interesting to make a comparison between data from figure 5.1 and table 5.1. It seems to be most likely that we have the majority of inhibiting influence from zinc and lead. Copper ions, as it was shown by us, have no significant influence on activity of immobilized AP [106]. For ppm concentration of heavy metal ions we have approximately same inhibiting activity for every one from five tested metals (fig 5.1), moreover no antagonistic or synergistic effects between different heavy metal ions has been reported for the *Chlorella vulgaris* based biosensor [3]. Thus we can suggest only simple adding of inhibition influences for zinc and lead ions.

There are some fluctuations of algal biosensor signal to wastewater samples (Fig 5.1). This can be explained by several reasons. Firstly, exact reproducibility of inhibition levels for enzymatic active live cell is not deeply investigated. In the case of biotesting of water with suspension of *Clorella vulgaris* it was reported similar fluctuations [93]. Thus, to overcome such obstacle, additional investigations of Chlorella *vulgaris* physiology have to be done. Secondly, fluctuations can be caused by interferences of real matrices with the biosensor. And thirdly, fluctuations clearly generated by the utilized in this study outmoded batch manual experimental setup. Such type of biosensors has to be utilized and stored during whole the period of measurements with the same thermostatic flow injection measuring cell. Thereby, this inconvenience can be resolved rather easily in the next step of investigations within the bounds of the project.

### 5.3.4. Bioassay responses after exposure to urban stormwater

Bioassays on microplate were carried out on free algae. Fig 5.2 shows phosphatase activity rate after 2 hours exposure of algal cells to different concentrations of stormwater samples.

A comparison between figures 5.1 and Fig 5.2 shows the same behavior between phosphatase activity of free cells and cells immobilized on conductometric biosensor: activation for lowest concentrations and inhibition for the highest. However, inhibition (or activation for low concentrations) rates are higher using biosensors than bioassays. It can be explained by the different ratios 'number of algal cells/toxicant elements' in both cases. Indeed on biosensors, low amounts of algae were immobilized compared to bioassays using free algae: for biosensors the ratio algae : toxicants is lower than for bioassays. As inhibition rate are inversely proportional to these ratios as a result, biosensors give higher inhibition rates and they seem to be more sensitive to detect the enzymatic activity modification.

Algal growth inhibition test was carried out on *Pseudokirchneriella sucapitata* according to AFNOR (NF T90-375 1998). Results showed very low toxicity of stormwater tested on algal growth (CE 50 was not reached with 80 % of stormwater in assay).

Then, changes in phosphatase activity reflect specific interactions between pollutants and algae and not a total attack of the cells.

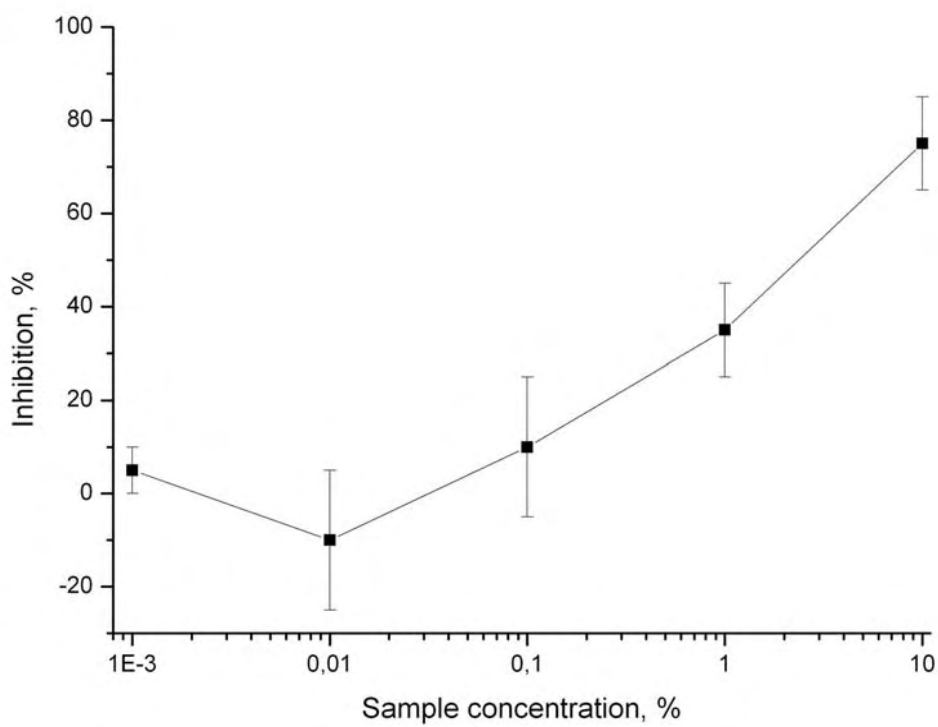

**Fig. 5.1.** *Evolution of algal phosphatase activity rate after exposure at different concentrations of urban stormwater detected with conductometric biosensor. Measurements were performed with 10 microns characteristic dimension platinum conductometric transducer deposited on silicon oxide substrate. Signal frequency and amplitude were 100 kHz and 10 mV respectively. Measurements were conducted in 10 mM Tris-nitrate buffer, pH 8.6 with 1mM Mg(NO₃)₂*

**Table 5.1.** *Heavy metal contain of tested urban stormwater*

| Zn ppm | Pb ppm | Cu ppm | Cr ppm | Ni ppm | Cd ppm | Fe ppm | Al ppm |
|--------|--------|--------|--------|--------|--------|--------|--------|
| 740 | 46.6 | 146 | 8.8 | 3.30 | 0.23 | 1.81 | 0.94 |

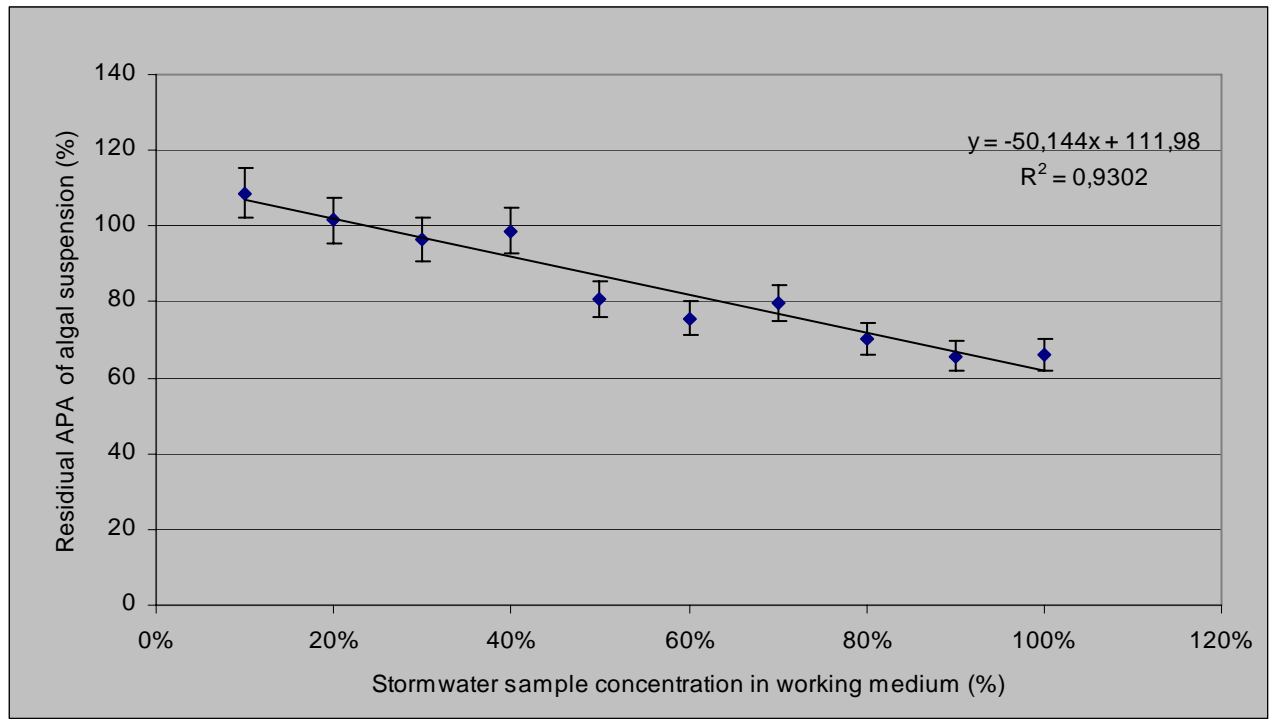

**Fig 5.2.** *Evolution of AP activity rate detected by 48 well microplate spectrofluorometer on free algae suspension after exposure to different dilutions of urban strormwater. Measurments were conducted in 0.1 M Tris-nitrare buffer, pH 8.6 with 1mM Mg(NO₃)₂*

### 5.4. Conclusions

The conductometric algal biosensor presented in this study seems to be successful as early warning system to identify the presence of pollutants as metal compounds.

In aquatic environment, algae are the first trophic level : any disturbances could be reported to upper levels. This is one of the main interest of this work since it gave the opportunity to follow the response of a living organism to pollutants. Contrary to chemical analysis, used to detect determined pollutants, algal cell biosensors can detect any stresses disturbing the organism metabolism.

Interferences of real matrices with these biosensors will have to be considered in the next step. After that, biosensors will be placed in the area under monitoring.

# Chapter 6

## Chapter 6. General conclusions and research perspectives

The fundamentals of conductometric sensors have been considered, taking into account their potential application in environmental monitoring. The options of modern microsystem techniques have been analyzed in terms of biosensor manufacture (transducer fabrication, methods of biomaterial immobilization, combination of several technologies in a single production cycle). A number of peculiarities of the most advanced microtechnologies used in production of conductometric biosensors have been presented, experimental measurement circuits demonstrated. The achievements in the field of development of enzyme conductometric biosensors have been described in detail, the prospect of their application evaluated. Some novel commercial systems for environmental monitoring based on biosensors have been exposed along with their practice.

At the initial stage of biosensor development, the key item is an appropriate alternative of the most effective transducer with regard to the best analytical characteristics and effective measurement procedure in real applications. Unfortunately, looking for optimal choice often is not based on complex analysis of the options of various systems of signal conversion, specific properties of transducers and bioselective materials. Therefore, comprehensive study of interactions between bioselective elements and transducers is extremely important for creating a general strategy of usage of transducers in biosensorics depending on the problem to be solved and the bioselective material chosen.

Concerning optimization of thin-film planar conductometric transducers, the priority line of various materials of the electrode sensitive part with regard to their usage in conductometric biosensors is proposed and is as follows: *platinum>gold>nickel>copper>chrome>titanium>aluminum*. The material of non-conducting substance is shown to be of no effect on the transducer sensibility, therefore, it can be chosen by its manufacturability and price. The characteristic dimensions of electrode sensitive parts are not a determinative parameter, and thus, a regular lithography technique can be utilized for electrode deposition in conductometric biosensors.

The results obtained were used by a number of institutions manufacturing conductometric transducers. The transducers produced at Kyiv Radioplant (Ukraine) and Institute of Chemo- and Biosensorics (Muenster, Germany) demonstrated the best characteristics and were recommended for development of conductometric biosensors.

The experimental devices for investigation of conductometric biosensors, as well as a portable measurement apparatus on its basis, were designed.

Furthermore the AP conductometric biosensors consisting of interdigitated gold electrodes and enzyme membranes formed on their sensitive parts have been used for an estimation of water pollution with heavy-metal ions. The application of immobilized AP has several benefits in comparison with free enzyme [82]: (1) thousands times lower consumption of immobilized enzyme [2, 84]; (2) reduction of interferences by the differential mode of operation ; (3) unnecessary pre-incubation; (4) less than 5 minutes rapid analysis procedure; (5) immediate inhibition of immobilized bovine AP by heavy metals ions is reversible, since it requires no re-activators like EDTA. Sensitivity of the biosensor to ionic strength is rather high as for all conductometric sensors and can be decreased with additional membranes [85]. The described conductometric biosensor can be used for analysis of water pollution by heavy-metal ions with detection limits about 2-100 ppm. Such concentration range is challenging for application in industry.

At the same time whole-cell conductometric biosensors consisting of interdigitated gold electrodes and algae entrapped in sol-gel membranes formed on their sensitive parts have been used for the assessment of water pollution with heavy metals ions.

Optimal algal concentrations in the membranes of the described biosensor were about $(30 \div 300)$ x $10^6$ cells/ml. Optimal pH of the biosensors working medium was 8.5.

And detection limits were about 1 ppb for $Cd^{2+}$, $Co^{2+}$, $Ni^{2+}$, $Pb^{2+}$ and 10 ppb for $Zn^{2+}$. The storage stability of the cells based biosensor was more than 40 days, while it was about 1 month for purified enzyme based sensor. In the view of reproducibility, at the present day, enzyme based biosensor is more reproducible than cells based one. We believe that reproducibility of algal sensor can be increased with some new perspective techniques and approaches. Incidentally sensitivity of our biosensors to ionic strength is rather high for all conductometric sensors and can be decreased with additional permselective membranes [98].

The data on development of different biosensors, including 2 proposed in this thesis, for heavy metals determination was summarized in the Table 6.1.

As it was mentioned in general introduction part, Celine Chauteau in her thesis has discovered the number of inconveniences related to utilization of GA and BSA for immobilization of algae on the transducer. In the same time, it was shown extremely challenging prospects for ppb level determination of heavy metals by such conductometric algal biosensor [3]. In the present work mentioned above inconveniences were overcome by the means of sol-gel technology for biomaterial immobilization. This support considerable reduce the detection limit for some heavy metals determination. In addition, sol gel entrapment makes possible lead analysis by this type of biosensors. Thus conductometric algal biosensor presented in this study seems to be successful core of early warning system to identify the presence of pollutants as heavy metal compounds. Such sensor devices based on AP are very low-cost analyzers. This economical characteristic is among the most important benefits of described biosensors.

As far as algae are the first trophic levels in aquatic environment: any disturbances could be reported to upper levels. This is one of the main interests of this work since it gave the opportunity to follow the response of a living organism to pollutants. Contrary to chemical analysis, used to detect determined pollutants, algal cell biosensors can detect any stresses disturbing the organism metabolism.

Interferences of real matrices with these biosensors will have to be considered in the next step. After that, biosensors can be placed in the area under monitoring.

Among the **perspectives** of this work there is a need to find the way of elimination of interference ions from bioselective membrane. There are our challenging preliminary results concerning utilization of NAFION membranes to protect the enzyme from anions, especially phosphate (up to 50 ppm $PO_4^{3-}$), which is known as inhibitor of AP and can be found in real samples.

It seems perspective to use other additional membranes to protect the enzyme and cells from bacterial contamination and to protect the conductometric transducer from interferential ionic force.

It's very important to use proposed biosensors with automatized flow-injection analysis systems. Multienzymatic sensor arrays with mathematical data treatment are also very desirable. However to apply multivariate analysis it has to be obtained data sets from different biosensors for example from set of enzymes as was showed in [107, 108]. Thus data sets received from 1-2 biosensors are insufficient to provide this type of statistical data treatment.

Thus with mentioned above recommendations described biosensor can be used as a part of early warning system for environmental monitoring.

**Table 6.1.** *The data on development of different biosensors and other bio-analytical systems for heavy metals determination*

| Bio sensitive element (1) | Transducer (2) | Immobilization method (3) | pH opt. (4) | Time of analys is (5) | Storage Stability (6) | Detection limit (7) | Remarks (8) | Ref. (9) |
|---|---|---|---|---|---|---|---|---|
| Urease | pH-ISFET | reticulation with BSA in GA vapors | ND | 30 min | ND | 1 ppm Cd, Pb, Cu, Co | Irreversible inhibition after 15 min contact with analyte | [109] |
| Urease | thick film interdigitated conductometric electrode | sol-gel entrapment | ND | 15 min | 7 days | 10 ppm Cd, Cu, Pb | Irreversible inhibition after 10 min contact with analyte | [74] |
| Urease | thin film planar interdigitated conductometric microelectrode | reticulation with BSA in GA vapors | ND | 30 min | 1 month | 0.5ppm Cd | Irreversible inhibition after 15 min contact with analyte | [2] |
| AP | potentiometry | free enzyme solution | ND | 10 min | ND | 5 ppm Cd 0.1 ppm Be | Reversible and immediate inhibition by Cd and activation by Ni, Co | [82] |
| Microalgae *Chlorella vulgaris* with AP activity | fiber optical detection | entrapment in micro fiber filter | ND | ND | ND | 10 ppb Cd, Pb | The system is based on massive optical detection system | [95] |
| Microalgae *Chlorella vulgaris* with AP activity | thin film planar interdigitated conductometric microelectrode | reticulation with BSA in GA vapors | ND | >1 hour | 21 days | 10 ppb Cd , Zn, Not sensitive to Pb | Irreversible inhibition after 1 hour contact with analyte and reversible inhibition after immediate contact. Insensitivity to Pb because of BSA utilization. | [3] |
| Microalgae *Chlorella vulgaris* with AP activity | 48 well microplate spectrofluorom eter | cells suspension | 10.5 | >4 hrs | > 1 month | 10 ppb Cd | The system is based on massive spectrofluorometer detection system | [3, 93] |
| AP | thin film planar interdigitated conductometric microelectrode | reticulation with BSA in GA vapors | 8.5 | < 5 min | 1 month | 2 ppm Cd, Zn, Ni, Co | This biosensor, presented in this study, can be used only for some industrial applications. DLs are insufficient for urban water control  (necessity of  as low DL as pbb level) | [70] |
| Microalgae *Chlorella vulgaris* wth AP activity | thin film planar interdigitated conductometric microelectrode | sol-gel entrapment | 8.5 | < 5 min | 40 days | 1 ppb Cd, Co, Ni, Pb; 10 ppb Zn | This biosensor, presented in this study, seems to be successful core of early warning system to identify the presence of pollutants as heavy metal compounds. | [87] |

**Titre en français: Développement des biocapteurs conductométriques à base de phosphatases alcalines pour le contrôle de la qualité de l'eau**

**Résumé en français**

Les recherches sont focalisées sur l'élaboration de dispositifs enzymatique micro-conducometriques permettant la détection des ions de métaux lourdes en phase aqueuse. La document comprend une **introduction générale** ; **un premier chapitre–la revue bibliographique**; un **deuxième chapitre** décrivant les fondements technologiques et méthodologiques des transducteurs conductométriques; un **troisième chapitre** examinant la possibilité de fabriquer le biocapteur conductométrique basé sur phosphatase alcaline bovine pour la détection des métaux lourds; un **quatrième chapitre** dévoue a la création du biocapteur conductométrique basé sur les technologie sol gel; la **dernière chapitre** examinant la possibilité d'application du biocapteur à cellules algales pour le contrôle des eaux urbaines; une **conclusion générale** sur les améliorations apportes a la technologie de biocapteurs pour contrôle de l'environnement

**Titre en anglais: Developpement of conductometric biosensors based on alkaline phosphatases for the water quality control**

**Résumé en anglais**

Researches are focused on the elaboration of enzymatic microconductometric device for heavy metal ions detection in water solutions. The manuscript includes: a **general introduction**; a **first chapter–bibliographic review**; a **second chapter** described the fundamentals of conductometric transducers; a **third chapter** examining the possibility to create and to optimize conductometric biosensor based on bovine alkaline phosphatase for heavy metals ions detection; the **fourth chapter** devoted to creation and optimization of conductometric biosensor based on alkaline phosphatase active microalgae and sol gel technology; the **last chapter** described application of the proposed algal biosensor for measurements of heavy metal ions toxicity of waste water; **general conclusions** stating the progresses achieved in the field of environmental monitoring.

**Titre en ukrainien: Розробка кондуктометричних біосенсорів на основі лужних фосфатаз для контролю якості води**

**Résumé en ukrainien**

Дослідження було сфокусовано на розробці ферментних мікрокондуктометричних біосенсорів для визначення вмісту важких металів у водних розчинах. Рукопис містить: **загальний вступ**, **літературний огляд** у **першій частині**; **друга частина** описує фундаментальні засади створення кондуктометричних перетворювачів; **третя частина** відображає створення та оптимізацію кондуктометричного біосенсора на основі бичачої лужної фосфатази для визначення вмісту іонів важких металів; **четверта частина** присвячена створенню та оптимізації кондуктометричних біосенсорів на основі мікроводорості із лужно фосфатазою активністю та золь-гель технології її іммобілізації; **остання частина** описує використання запропонованого біосенсора для оцінки вмісту важких металів в стічній воді. Роботу завершує **загальне заключення**, у якому підсумовано основні досягнуті результати та намічено перспективи робіт у даному напрямку.